\documentclass[twocolumn]{aastex701}

\usepackage{multirow}
\usepackage{color}
\usepackage[mathlines]{lineno}

\linenumbers
\usepackage{capt-of}
\usepackage{longtable}
\usepackage{amsmath,amsfonts,amssymb}
\usepackage{xcolor}
\usepackage{natbib}
\usepackage{chngcntr}
\usepackage{enumitem}
\setlist{parsep=0pt,listparindent=\parindent}
\usepackage{xtab}
\usepackage{tabularx} 
\usepackage{xltabular} 
\usepackage{soul}
\usepackage{array}
\usepackage{physics}
\usepackage{booktabs}
\usepackage{rotating}

\usepackage{mwe}
\usepackage{rotating}
\usepackage{afterpage}

\usepackage{colortbl}
\usepackage{lipsum}

    {\global\pdfpageattr\expandafter{\the\pdfpageattr/Rotate 90}}%
    {\clearpage\pagebreak[4]\global\pdfpageattr\expandafter{\the\pdfpageattr/Rotate 0}}%

\usepackage[utf8]{inputenc}
\usepackage[T1]{fontenc}   
\usepackage{url}
\usepackage{nicefrac}
\usepackage{microtype} 
\usepackage{graphicx}
\definecolor{dkgreen}{rgb}{0,0.6,0}
\definecolor{dkdkgreen}{rgb}{0,0.3,0}
\definecolor{gray}{rgb}{0.5,0.5,0.5}
\definecolor{mauve}{rgb}{0.58,0,0.82}

\DeclareGraphicsExtensions{.png,.pdf}

\usepackage{upquote}
\usepackage{appendix}

\usepackage{listings}

\usepackage{dsfont}

\newcommand{\mbf}[1]{\mathbf{#1}}

\newcommand{\R}{\mathbb{R}}
\newcommand{\C}{\mathbb{C}}
\newcommand{\N}{\mathbb{N}}
\newcommand{\Z}{\mathbb{Z}}
\newcommand{\F}{\mathbb{F}}
\newcommand{\Q}{\mathbb{Q}}

\newcommand{\x}{\mbf{x}}

\begin{document}

\title{\emph{Mantis Shrimp:} Exploring Photometric Band Utilization in Computer Vision Networks for Photometric Redshift Estimation}

\author[0000-0003-2348-483X]{Andrew W. Engel}
\affiliation{National Security Directorate, Pacific Northwest National Laboratory (PNNL),
Richland, WA, USA}
\affiliation{Department of Physics, Ohio State University, Columbus, OH, USA}
\affiliation{Center for Cosmology and AstroParticle Physics (CCAPP),
Ohio State University, Columbus, OH, USA}
\email{engel.250@osu.edu}

\author[0000-0002-7392-3637]{Nell Byler}
\affiliation{National Security Directorate, Pacific Northwest National Laboratory (PNNL), Richland, WA, USA}
\email{eleanorbyler@gmail.com}

\author[0000-0001-6116-2984]{Adam Tsou}
\affiliation{Department of Applied Mathematics and Statistics, Johns Hopkins University, Baltimore, MD, USA}
\email{atsou2@jh.edu}

\author[0000-0001-6022-0484]{Gautham Narayan}
\affiliation{Department of Astronomy, University of Illinois at Urbana-Champaign, Urbana, IL, USA}
\affiliation{Center for AstroPhysical Surveys, National Center for Supercomputing Applications, Urbana, IL, USA}
\affiliation{Illinois Center for Advanced Studies of the Universe, University of Illinois, Urbana, IL, USA}
\email{gsn@illinois.edu}

\author[0000-0003-3932-9105]{Emmanuel Bonilla}
\affiliation{Research Computing, Pacific Northwest National Laboratory (PNNL),
Richland, WA, USA}
\email{emmanuel.bonilla@pnnl.gov}
\author[0000-0002-9279-3006]{Ian Smith}
\affiliation{Research Computing, Pacific Northwest National Laboratory (PNNL),
Richland, WA, USA}
\email{ian.smith@pnnl.gov}

\submitjournal{ApJ}

\correspondingauthor{Andrew William Engel}
\email{engel.250@osu.edu}



\begin{abstract}
We present \texttt{Mantis Shrimp}, a multi-survey deep learning model for photometric redshift estimation that fuses ultra-violet (GALEX), optical (PanSTARRS), and infrared (UnWISE) imagery. Machine learning is now an established approach for photometric redshift estimation, with generally acknowledged higher performance in areas with a high density of spectroscopically identified galaxies over template-based methods. Multiple works have shown that image-based convolutional neural networks can outperform tabular-based color/magnitude models. In comparison to tabular models, image models have additional design complexities: it is largely unknown how to fuse inputs from different instruments which have different resolutions or noise properties. The \texttt{Mantis Shrimp} model estimates the conditional density estimate of redshift using cutout images. The density estimates are well calibrated and the point estimates perform well in the distribution of available spectroscopically confirmed galaxies with (bias = 1e-2), scatter (NMAD = 2.44e-2) and catastrophic outlier rate ($\boldsymbol{\eta_{>0.15}}=4.51$\%). We find that early fusion approaches (e.g., resampling and stacking images from different instruments) match the performance of late fusion approaches (e.g., concatenating latent space representations), so that the design choice ultimately is left to the user. Finally, we study how the model learns to use information across bands, finding evidence that our model successfully incorporates information from all surveys. \added{The applicability of our model to the analysis of large populations of galaxies is limited by the speed and ease of downloading and preparing cutouts from external servers;} however, our model could be useful in smaller studies such as in generating priors over redshift for stellar population synthesis.
\end{abstract}

\section{Introduction}
Measuring redshifts to distant galaxies is fundamental to extragalactic astronomy and cosmology research. The gold standard for redshift measurements is spectroscopic observations, but spectroscopy is expensive and time-consuming compared to photometric observations. Algorithms that use photometry to estimate redshift have emerged as an alternative to match the scale of on-going \citep{KIDS_plus_VIKING_BPZ,DES_Photometric_Redshifts,RedshiftsCFHS,eulid} and future \citep{LSST_ScienceBook2,RomanDoc} photometric surveys. Obtaining more accurate photometric redshifts is in fact necessary to complete these survey's science objectives as errors associated with them are still the leading term of uncertainty in multiple cosmological analyses, including weak lensing \citep{WeakLensingReview18}. See \citet{NewmanReview} for a recent review. 

There are two historical strategies to estimate redshift from photometry: template fitting approaches and empirical approaches. In template fitting approaches, a library of galaxy spectra is convolved with observatory photometric bandpass filters to measure the expected observed flux as a function of redshift \citep{FirstTemplateMethod1996,LephareTemplate,BPZ,HyperZTemplate,EasyTemplateLibraries,ZebraTemplate}. These expected fluxes are compared with the observed flux to constrain the correct galaxy spectral model and its redshift. In contrast, empirical approaches use machine learning algorithms to learn the mapping between observed photometry and redshift from a labeled dataset collected from spectroscopic surveys. It is generally acknowledged that in areas of high label density (i.e. within the targeting criteria of the major spectroscopic surveys and consequently nearby in redshift), these empirical approaches tend to outperform template fitting models \citep{EuclidDataChallengePhotoZ,NewmanReview}. In this work, we will focus our attention onto empirical approaches.

\label{sec:related_work}
\label{sec:related_work_PZ}

Previous empirical photometric redshift works have investigated many kinds of machine learning algorithms,
\citep{ANNz2_and4000A_break_Affects_PZ,ANNzCollister2003,SDSS_PhotoZ_GaussianProcesses,Beck2016LLNSDSS,Zhou_Photoz_DECaLS_DR9,tarrio}, but recent works have utilized computer vision models (e.g., Convolutional Neural Networks or CNNs) to estimate the redshift directly from cutouts, rather than from derived features available in catalogs \citep{Pasquet2019,Dey2021Capsule,Hayat_2021_selfsupervised,CNN_multichannel_PZ_DIsanto2018,FirstCNNHoyle2015,CNNSDSSLargeHenghes2021,AstroCLIP}. In these works, computer vision networks show replicable improved performance across the range of community accepted point-estimate metrics (summarized in section~\ref{sec:metrics}) in the SDSS main galaxy sample \citep{SDSS_MGS} over tabular models \citep{NewmanReview}. In particular, \citet{Pasquet2019}, hereafter P19, is an influential work that first showed major improvements over the official SDSS photometric redshift algorithm \citep{Beck2016LLNSDSS}. Computer vision models are believed to exceed tabular performance because they bypass manual feature extraction, allowing artificial intelligence to learn directly from image pixels and the full two-dimensional variation in galaxy surface brightness, size, and morphology \citep{Dey2021Capsule}. However, it has not been shown whether computer vision models still outperform tabular models in a higher redshift regime, which is what we will explore in this work.


Prior works have also demonstrated that deep learning models can enhance photometric redshift estimation by combining observations across multiple surveys, but have been limited to exploring tabular models \citep[e.g.,][]{WISE-PS1-STRMBeck22, KIDS_plus_VIKING_BPZ, DESI_redshifts_Zhou21, CircleZ_AGN_LegacySurvey_grizW1through4}. Notably, \citet{WISE-PS1-STRMBeck22} (hereafter B22) showed that integrating WISE with the PanSTARRS catalog improves upon their previous work using PanSTARRS alone \citep{Beck21PS1STRM}. This study extends B22's work by incorporating GALEX, PanSTARRS, and WISE photometry and utilizing image cutouts instead of tabular data.

Our choice to fuse cutouts from various surveys forms a unique challenge that has not been addressed in prior works: what strategies can be used from the computer vision literature to combine images of varying pixel resolutions, depths, from different photo-reduction pipelines? Furthermore, how do these different strategies compare against each other in both model behavior and performance? Specifically, we investigate two such strategies: we can resample images of different pixel resolution to a common pixel scale (early fusion) or keep the native resolution and combine a latent feature vector specific to each survey together (late fusion). It can be difficult to ascertain whether a multi-modal model is leveraging all the available information across each photometric band \citep{UniModal_Collapse}. One aim of this work is to evaluate the trade-offs and behavioral differences between these fusion approaches in the context of a model for photometric redshift estimation. For a review of modality fusion in computer vision, see \citet{MultiModalityReview}.

One unique advantage that computer vision models hold over tabular models are that computer vision methods do not rely upon a detection of an object in each survey to make a valid join between catalogs. For example, if a source is detected in PanSTARRS, but is not detected in WISE, a catalog join between the two surveys would possibly match to the nearest but incorrect source in WISE, or simply have missing flux values. Computer vision models bypass photo-reduction pipelines to be run at arbitrary user-given coordinates. This is an additional freedom that we leave exposed to users, something that we call a "forced photo-z". This freedom comes with additional responsibilities that we leave to the user to evaluate the context of our model's predictions in terms of blending, low signal-to-noise, etc, that we do not consider in this work.

\label{sec:related_work_Interp}
Given the increasing prominence of computer vision in the physical sciences plus the rise of large language models whose ability to be partners in the scientific process are currently being evaluated \citep{Mephisto_LLM_YuanSen,AstroLLAMA_YuanSen}, an emerging topic of research is in defining the role domain scientists can play in the evaluation of AI models. In this work, we qualitatively explore the utility of each photometric band to the importance we measure each band plays for our AI model's photometric redshift estimate. We employ Shapley Values to measure feature importance \citep{ShapleyOG,ShapleyRepopularizedforNN}, which have become a popular tool in AI interpretability. Previous works in the field have attempted to explain the performance of photometric redshift algorithms through feature importance \citep{FeatureImportanceDIsanto}, but did not compare to the expected behavior of the model given domain knowledge. Additionally, prior works \citep{InterpretabilityCosmologyNtampaka2022} investigated the use of Shapley value interpretability measures for cosmology, but not specifically for photometric redshift estimation. This specific framing of comparing the expected behavior given domain knowledge to the behavior we infer from Shapley values is unexplored in prior works. 


To summarize, we present \texttt{Mantis Shrimp}, a multi-modal CNN for the estimation of photometric redshifts and the associated conditional density estimates (CDEs). We will use a convolutional neural network to combine images from the GALEX, PanSTARRS, and WISE surveys. We match image cutouts from each survey centered on galaxy coordinates with matching spectroscopic redshift labels, compiled from a list of major spectroscopic surveys. Our model predicts a completely non-parametric CDE of redshift given the cutouts, which we can use to make point estimates and perform uncertainty quantification. In contrast to previous studies, our model learns to extract features directly from images from multiple surveys. We make our unique multi-modal galaxy cutout dataset available to download via \added{Pacific Northwest National Laboratory's} (PNNL) DataHub\footnote{\href{https://data.pnnl.gov/group/nodes/dataset/33966}{\texttt{Mantis Shrimp} Dataset on PNNL DataHub}}.
We make our model training and data collection pipeline available as an open-source GitHub repository\footnote{\href{https://github.com/pnnl/MantisShrimp/tree/v1.0.0}{\texttt{Mantis Shrimp} GitHub Repository}}.
We host a web-app\footnote{\href{https://mantisshrimp.pnnl.gov}{\texttt{Mantis Shrimp} Web-app}} where users may evaluate our model by simply uploading a coordinate. We hope that the ease-of-use of our web-app and data already formatted for deep learning applications is of additional value to the community.


\section{Background}
\label{sec:Background}

\subsection{The Photometric Redshift Conditional Density Estimation Task}
\label{sec:neural_networks_definition}
Our goal is to model a density estimate of the spectroscopic redshift  conditioned on the observed cutout data (and implicitly our training). The advantage of modeling the density rather than just providing point-estimates is that we can compute confidence regions and follow-on analyses can leverage the full complexity of our non-parametric density estimate (e.g., using our density estimates as a prior for redshift in stellar population synthesis modeling \citep{Prospector}). We formally introduce the photometric redshift density estimation task in the following paragraph. See \citet{ConditionalDensityEstimationToolboxDalmasso2019} for another formal introduction to conditional density estimation for this task.

While the \texttt{Mantis Shrimp} dataset is composed of paired cutout images and scalar spectroscopic redshifts, the photometric redshift conditional density estimation task is more easily framed as a supervised classification task, and so in what follows we will describe a dataset formed from cutouts and class labels. We construct classes by binning redshifts into $C=400$ classes. Given a choice of maximum redshift $Z_\text{max} = 1.6$, the bin-width is $\delta_z = \frac{Z_\text{max}}{C}$. In what follows let $\mathcal{E}_C = \{\mbf{e}_1,\mbf{e}_2,\ldots,\mbf{e}_C\}$, where $\mbf{e}_i$ is the $i$-th standard normal basis vector in $\mathbb{R}^C$, or equivalently in regular machine learning parlance is a one-hot encoded label of class $i$.

Let $D$ be a labeled dataset of length $N$ composed of individual data-label pairs drawn from the population of high confidence spectroscopically identified galaxies, $(x_i, c_i) \subseteq (\mathcal{X},\mathcal{E}_C)$ with data in the space of hyperspectral images, $\mathcal{X} \in \mathbb{R}^{B \times H \times W}$, where $B$ is the number of photometric bands ($B$=9), \added{H is the height, and W is the width,} such that $D=\{(x_1, c_1),(x_2, c_2),\ldots,(x_N, c_N)\}$. The class label $c_i$ is determined by the true spectroscopic redshift $z_i$ as $c_i = \mbf{e}_j \; | \; j = \text{floor} \lfloor \frac{z_i}{\delta_z} \rfloor$. Our neural network function $F$ is a parameterized and learnable mapping from galaxy hyperspectral image cutouts to a redshift latent vector, $F: \mathcal{X} \to \mathcal{Y}$ with $\mathcal{Y} \subseteq \Delta^C$, where $\Delta^C$ is the C-dimensional unit simplex. Let $\theta$ be the vector of learnable parameters. The output of the network is a vector where each element $y^j_i$ is approximately the probability that the sample has true redshift in a small redshift bin, $y^j_i \approx P(\delta_z \cdot (j-1) < z_i \leq \delta_z \cdot j)$.  We will evaluate whether the values of $y^j_i$ are well-calibrated to reflect the true probability empirically. The weights $\theta$ are learned via gradient descent to minimize the cross entropy loss evaluated between $y_i$ and $c_i$. Using this interpretation of $y_i$, we define our point estimate of redshift, $\hat{z}_i$, as the expected value $\hat{z}_i = \mathbb{E}(y_i)$. 

As a final note, $y_i$ as defined above is formally a probability mass function; however it is common practice to interpolate values on $y_i$ to define a probability density function that can be evaluated at arbitrary redshift along the interval $(0,\text{Z}_{\text{max}})$ \citep{Dey_ConditionalCalibration2022_CalPIT}. This requires a re-normalization to satisfy $\int y_i \mathrm{d}z = 1$. We will continue to refer to $y_i$ as a conditional density estimate and in-fact report values from our web-app with this re-normalization applied by default. 

\subsection{Galactic Radiation and Photometric Redshifts}
\label{sec:physics}

As part of our analysis into the behavior of our neural network models, we will investigate how the models use information from each photometric band and whether that use agrees with our expectations given astronomical domain knowledge. To make this assessment, we will briefly introduce some background on galactic radiation. To understand photometry, we introduce a model of observed flux in a band as the convolution of a galaxy's observed spectral energy density (SED) and the transmission curve of that band.

Galaxy spectra are the synthesis of all radiative and absorptive processes in a galaxy, dominated by the stellar population present in the galaxy. The result of this synthesis is a roughly bell-shaped SED, with upper tail at a rest frame wavelength of 912\AA, (the Lyman limit, caused by the ionization energy of the neutral hydrogen atom), or 4000\AA~for galaxies particularly bereft of young hot UV emitting stars \citep{4000Break_explanation}. The lower tail of the spectra occurs from 1$\mu$m - 5$\mu$m, and unlike the upper tail, this IR feature is common to many galaxies independent of Hubble type or star formation history \citep{Mannucci2001NIRsame}. 

We can model the total flux density observed, $\Phi$ [Jy], from an object in any particular photometric band with transmission curve $R(\lambda)$ [unitless] and galaxy's SED $F_\lambda(\lambda)$ [Jy m$^{-1}$ s$^{-1}$] as:
\begin{equation}
\Phi = \frac{\int_{\lambda_a}^{\lambda_b} F_{\lambda}(\lambda) R(\lambda) \lambda \mathrm{d}\lambda}{\int_{\lambda_a}^{\lambda_b}  c \frac{1}{\lambda} R(\lambda) \mathrm{d}\lambda}
\label{eq:SED}
\end{equation}
The observed SED redshifts as $\lambda_{obs} = \lambda_{emit}(1+z)$, but the photometric band transmission curves stay fixed in our observation frame.

\begin{figure*}[!ht]
    \centering
    \includegraphics[scale=0.5, trim=2mm 2mm 2mm 2mm, clip]{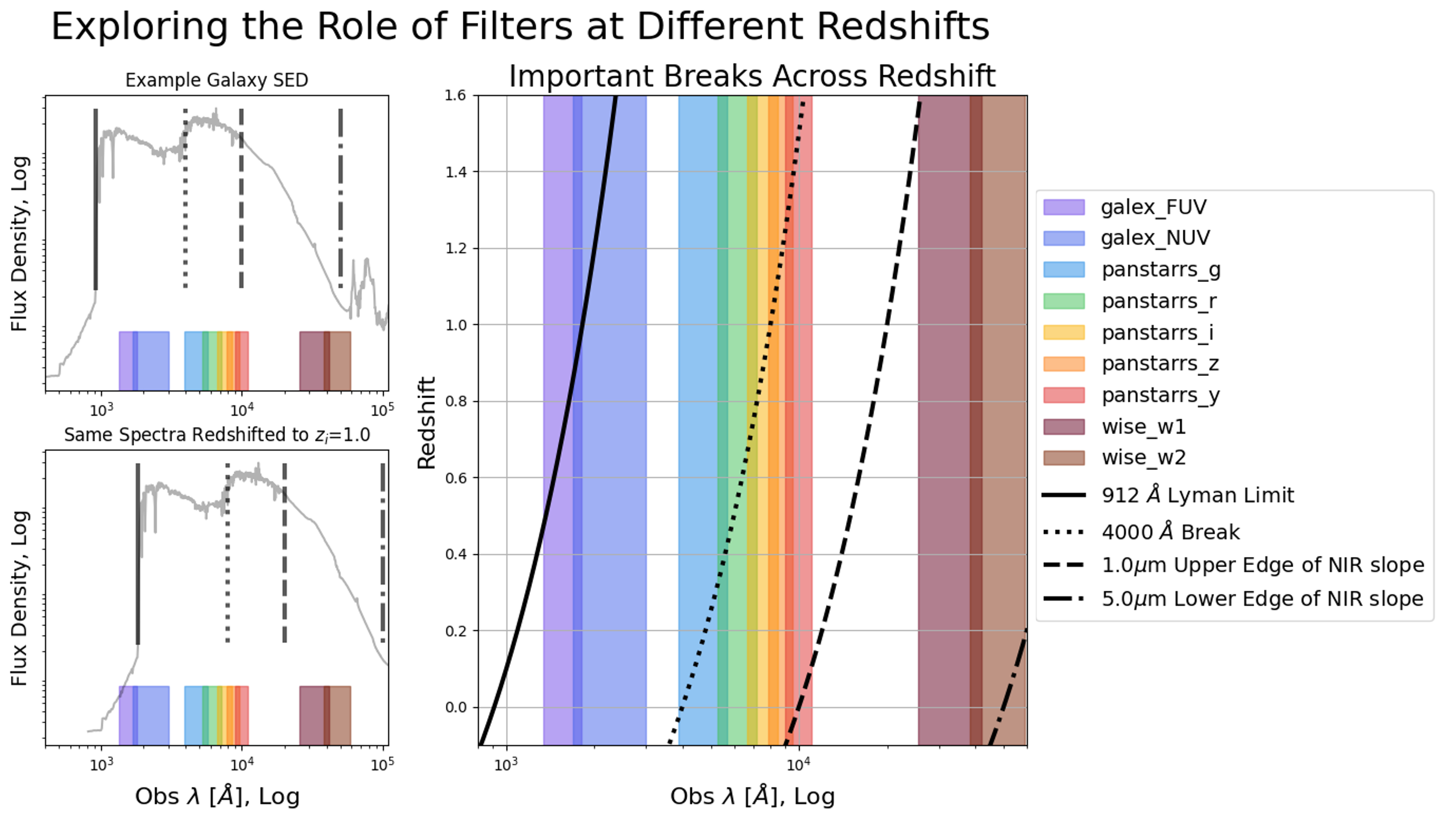}
    \caption{\textbf{Top Left.} A representative galaxy SED (taken from the atlas in \citet{Brown14_galTemplates}) is plot at that galaxy's rest-frame, with the photometric filters used in this work overlaid. Additionally, important breaks in the galaxy SED are identified with vertical black lines. \textbf{Lower left.} We plot the same SED except redshifted to a value of 1.0 to visually represent how light incoming from a galaxy appears in our observation frame. The important breaks/phenomena have shifted with the spectra and now coincide in different parts of photometric band-space. \textbf{Right.} Finally, we plot important features generally common to galaxy SED's characteristic wavelength as a function of redshift.}
    \label{fig:physics2}
\end{figure*}

We visualize an example galaxy SED, labeled with the relevant major features and photometric bands used in this work, in Figure~\ref{fig:physics2}. As an SED becomes redshifted, energy density leaves the UV/optical bands and moves into the IR bands. This overall shift can explain why the $r - W_1$ color is known to be a good predictor of redshift, and in fact is used in DESI spectroscopic targeting \citep{DESI_LRG}. Individual features such as the 4000\AA~break move through filters successively. At low redshifts, the 4000\AA~break should be important as the break transitions through the $g$ band and into the $r$ band, making the color $g - r$ predictive of redshift, until about $z=0.4$, when the  4000\AA~break transitions out of the $g$ band \citep{4000Abreak_affectsPZ}. Since this intuition ties the performance of the model to an object's redshift and specific photometric bands, we ask whether the same intuition could be applied to understand our model behavior. 

\subsection{Shapley Values For Band Importance}
\label{sec:shapley_values_definition}
To analyze how the model uses different photometric bands over the range of redshifts we will perform a Shapley value analysis. Shapley values have seen a resurgence in popularity as a tool for neural network explainability \citep{ShapleyRepopularizedforNN}. As intuition, Shapley values were originally introduced to measure how ``valuable'' a player is to a team in an N-person game \citep{ShapleyOG}; in analogy, we will use Shapley values to measure how important the flux from the targeted galaxy in each photometric band is to the point estimation of redshift. Shapley values are computed by ablating features then observing the effect of that ablation on the point estimate of redshift. This comparison is performed across all possible permutations of feature-ablation to understand the synergy between features. In our implementation we will define the flux of the target galaxy in each band as the feature that we will ablate. We describe the specifics of our implementation in Section~\ref{sec:shapley_methods}, but for now we formally introduce Shapley values.  

Assume we have a hyperspectral image dataset, with samples $x_i \in \mathcal{X}$. Recall that $B$ is the total number of bands, and let the $j$-th band of $x_i$ be $x_i^j$. As mentioned above, Shapley values ablate features, and so for clarity let there be a baseline image $b_i \in \mathcal{B} \subseteq \mathbb{R}^{B \times H \times W}$ which contains the ``feature ablated'' images found by a feature ablation function, $\mathcal{A}: \mathcal{X}^j \to \mathcal{B}^j$, which maps the original observed band image to the baseline, $\mathcal{A}(x_i^j) = b_i^j$.

Next, we define $S$ to be a subset of the channel indices, $S \subseteq \{1,\ldots,B\}$, where the set $S$ contains the indices of bands to be left unmodified in a forward pass of the model. Let $\mathcal{P}(S)$ be the power-set of $S$, such that $\mathcal{P}(S) = \{\{\emptyset\},\{1\},\{1,2\},\ldots,\{1,\ldots,B\},\ldots,\{2\},\{2,3\},\ldots,\{B\}\}$. Furthermore, we will slightly abuse notation and allow $\mathcal{P}(S) / \{j\}$ to mean the powerset of $S$ but with subsets that included the integer $j$ having been removed. We can then define a feature replacement function, $\mathcal{H}: (\mathcal{X},\mathcal{P}(S)) \to \mathcal{X} \cup \mathcal{B}$, which replaces the bands whose indices are not in S with the baseline image bands, $\mathcal{H}(x_i,S) = \Tilde{x}_i \; | \; \Tilde{x}_i^j = b_i^j \, \forall \, j \notin S$. We define the value of including bands, $\mathcal{V}(S)$, to be equal to the model's point prediction of redshift on the modified $\Tilde{x}_i$ input. For clarity, $\mathcal{V}(S) = \mathbb{E}[F(\mathcal{H}(x_i,S),\; \theta)]$. 

The Shapely value, $\mathcal{S}$, for the $j$-th band measured on $x_i$ is defined:
\begin{equation}
\small
  \mathcal{S}_j(x_i):=\sum_{S \in \mathcal{P}(S) / \{j\}}  (\frac{|S|!(B-|S|-1)!}{B!}) \times (\mathcal{V}(S \cup {j}) - \mathcal{V}(S)).
  \label{eq:shapleyvalue}
\end{equation}
That is, the Shapley value is a weighted sum where the left term in the product of \eqref{eq:shapleyvalue} weighs images with more features ablated less than images with less total features ablated. The right term is a simple difference in the point estimate of redshift measured on a modified image compared to including or ablating the $j$-th band's target galaxy flux. We note that the Shapley value can be positive, negative, or zero indicating whether including the band information is increasing (positive) or decreasing (negative) the point estimate of redshift. To measure the importance of photometric bands for our entire dataset across redshift, we will average the Shapley value across samples in bins of redshift to understand the change in band importance with redshift.  

A slight limitation of Shapley values are that because they can be negative it can be difficult to understand the relative importance of each band. As a remedy, we will also employ the MM-SHAP value \citep{WhichInputMultiModal}, which is simply a normalization of the absolute magnitude of the Shapley values across each band: 
\begin{equation}
\text{MM-SHAP}_j(x_i) = \frac{|\mathcal{S}_j(x_i)|}{\sum_j^F |\mathcal{S}_j(x_i)|},
\end{equation}
and will similarly apply binning our samples in redshift to understand trends in band importance across redshift. 

To summarize, we will use the Shapley values to roughly answer "\textit{how is the presence of this band impacting our photo-z estimate?}" and MM-SHAP to answer "\textit{what is the relative importance of each band?}" It is worth re-emphasizing that both the Shapley value and MM-SHAP are defined specific to a choice of modified datapoint $\Tilde{x}_i$. The choice of how to represent null contribution from a photometric band is therefore critical to our interpretation of the model and is often non-trivial \citep{DistilFeatureBaselines}. For example, a reasonable but naive choice would be to replace the band with all zeros, or replace with Gaussian noise sampled from the image. We will instead use domain-specific software to remove the flux specific to the targeted central galaxy in the cutout. We discuss our domain specific methodology in Sec~\ref{sec:shapley_methods}.

\section{Data}
\label{sec:data}
\label{sec:data_main}
\begin{figure*}[!ht]
    \centering
    \includegraphics[scale=0.6]{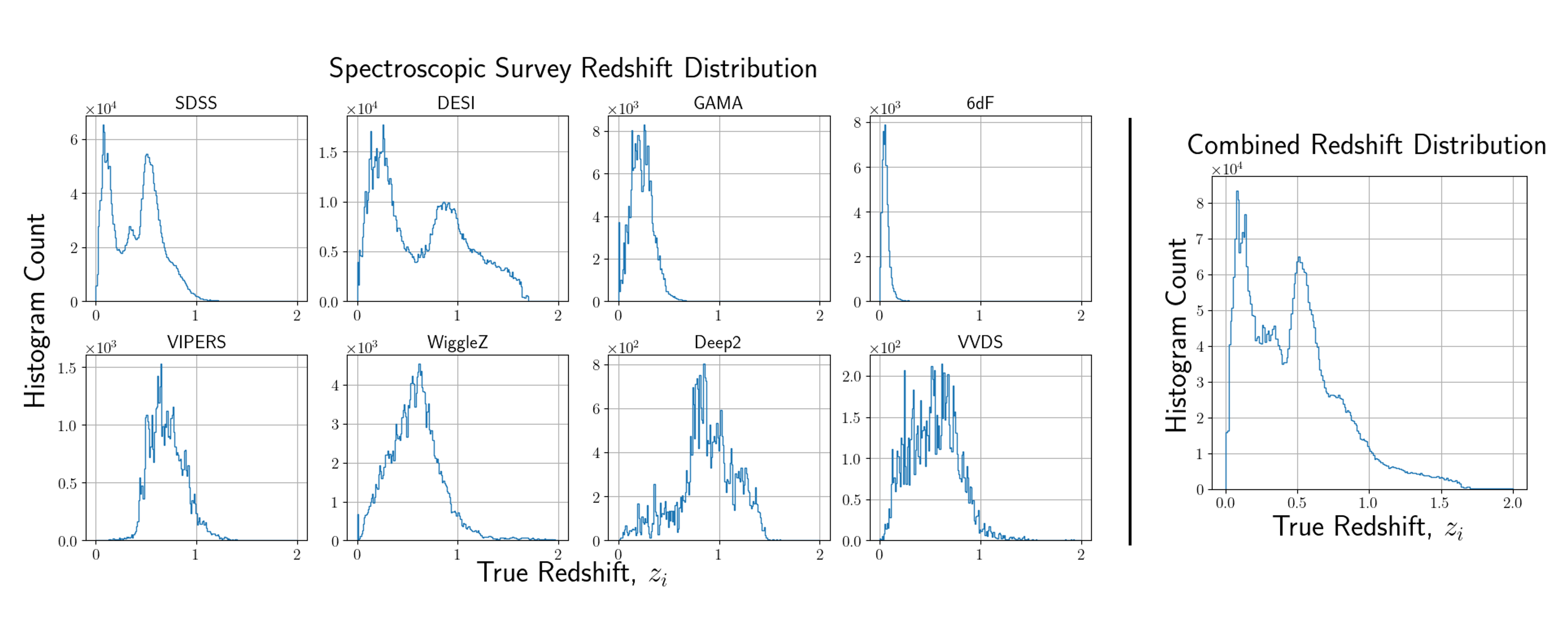}
    \caption{\small\textbf{Left} Distribution of targets in redshift visualized by individual survey. \textbf{Right} Combined distribution of redshift labels. Notice that the shape of the combined distribution is dominated by SDSS and DESI.}
    \label{fig:spectroscopicdist}
    \label{fig:combined_spectroscopicdist}
\end{figure*}

We provide a brief overview of the \texttt{Mantis Shrimp} dataset in this introductory section then go into greater detail in the following subsections. The data consists of $N=4.4\times10^6$ 9-band image cutouts of galaxies paired with ground-truth redshifts compiled from a diverse set of spectroscopic surveys \citep{SDSS_four,DESI_EDAspectroscopy, DEEP2, DEEP3, GAMA_DR2, VVDS_final_release, VIPERS_final, 6dFGS_final, WiggleZ_final}. Although spectroscopic redshifts are the most secure method of identifying redshift, we employ cuts based on provided spectroscopic survey reported quality to ensure a high-confidence sample of labels for our model, following a similar methodology as in B22. \added{In Figure~\ref{fig:spectroscopicdist} we visualize the absolute and relative distribution of redshifts from each component survey. In Figure~\ref{fig:ExplorationOptical} we visualize the distribution of our sample of spectroscopically confirmed galaxies in magnitude and color space compared to a random sample of PanSTARRS and DESI Legacy Survey's Galaxies.}

\begin{figure*}[!ht]
    \centering
    \includegraphics[scale=0.29]{./ExplorationOptical.png}
    \caption{\added{We explore the distribution of the \texttt{Mantis Shrimp} sample in the context of colors and magnitudes of a random selection of galaxies from the PanSTARRS forced mean photometry table and the DESI Legacy Survey DR10 Tractor model photometry table. Note that the histograms along the diagonal are each individually normalized for each population and in log scale. This log scale highlights the tails of the distribution which better demonstrates the coverage of our sample in context with a random sample. The plots confirm that the set of spectroscopically confirmed galaxies are heavily biased towards bright objects, which should be expected due to the relative ease of acquiring high quality spectra from bright objects. While there appears to be high coverage of the spectroscopic sample in magnitude space up to about the 5$\sigma$ limit of the PanSTARRS galactic photometry, the color-magnitude plots reveal that spectroscopic galaxies are also highly concentrated in color space. We return to this figure in our discussion as an important motivation for future work in characterizing the completeness of spectroscopic galaxies.}}
    \label{fig:ExplorationOptical}
\end{figure*}

\begin{figure*}[!ht]
    \centering
    \includegraphics[scale=0.5]{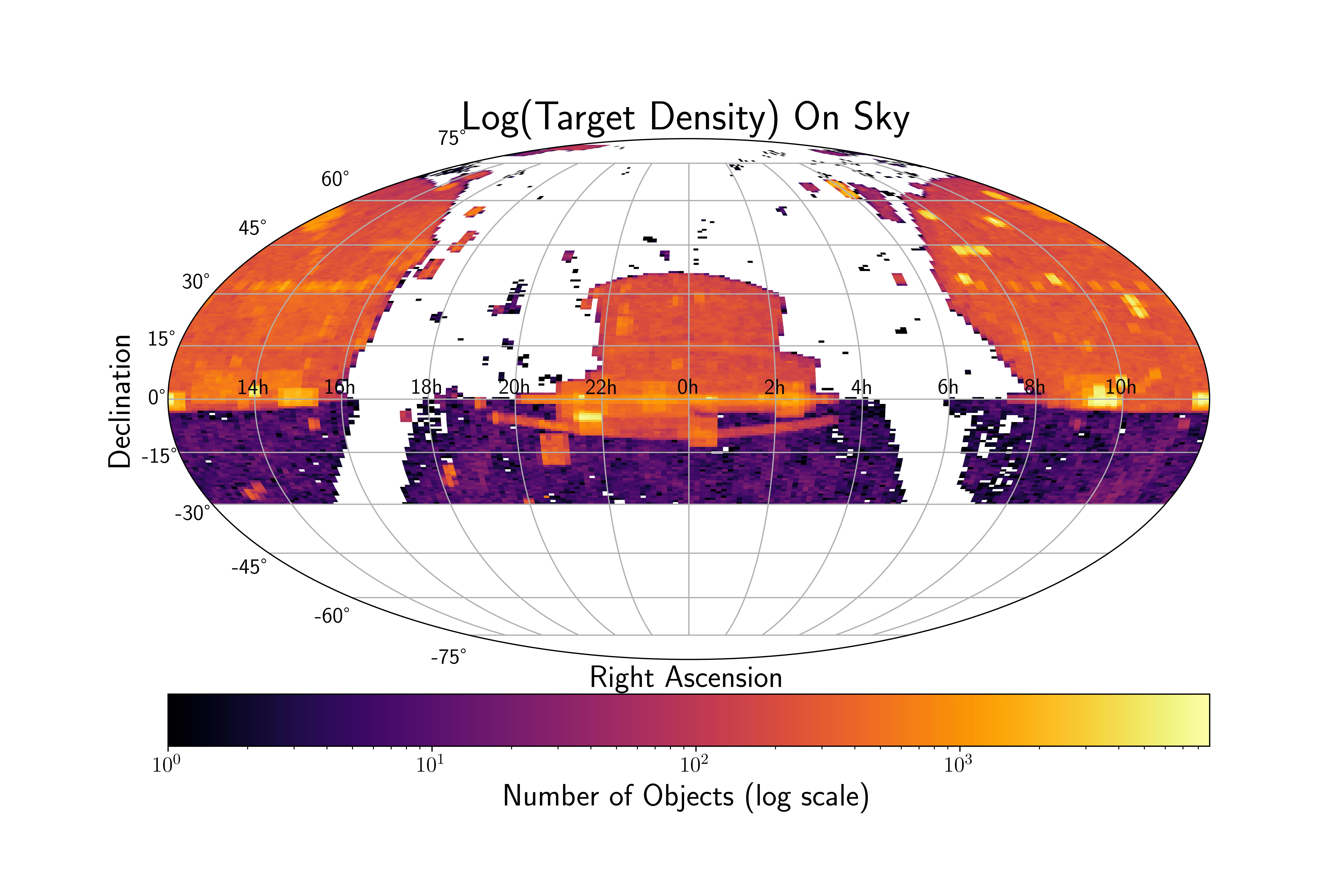}
    \caption{Distribution of targets in the sky, plot as astrometric position on the night sky (RA, DEC). Our spectroscopic sample is dominated by the SDSS and DESI footprint in the northern sky, with some deep drilling wells from spectroscopic surveys showing as particularly bright spots. The southern sky south of DEC=$-30^\circ$ is excluded due to the PanSTARRS footprint. The sparse region of the northern sky spans the Milky Way.}
    \label{fig:sky_pos}
\end{figure*}

To construct our image inputs, we query public APIs of image cutout servers for each of the three instruments, centered on the coordinates provided by the spectroscopic surveys. We include 2 UV bands from GALEX \citep{GalexSurveyPaper}, 5 optical bands from PanSTARRS \citep{PSSurveyPaper}, and 2 IR bands from UnWISE \citep{WISE_og,UnWISE_og}. We describe in greater detail the collection of photometric data in appendix~\ref{appendix:photometry_download_and_description}, while image pre-processing and our augmentation pipeline are detailed below. To summarize: we ensure our images are on a quasi-logarithmic flux scale \citep{Luptitude}, then map all pixels with flux less than zero to zero. It has been shown that quasi-logarithmic scales for astronomical images work well in convolutional neural networks for segmentation tasks \citep{LuptonScalingInResNets_Colin_Grant_Patrick_detectron2}. Where our pipeline encounters NaNs (typically caused by bright-source masks in the PanSTARRS cutouts), we replace with zero. We show the on sky distribution of labeled galaxies in Figure~\ref{fig:sky_pos}. \added{For the analysis requiring it, we perform cross match queries onto the PanSTARRs DR2 catalog \citep{PanStarrsDR2CatalogMAST} and GALEX catalog \citep{GalexCatalogMAST} hosted by MAST, and the UnWISE and DESI Legacy Survey catalogs.}

\subsection{Spectroscopic Survey Compilation}
\label{sec:spectroscopic_survey_compilation_mainbody}

\begin{table*}[!ht]
\caption{\small\textbf{Spectroscopic Band Table} $N_\text{avail}$ is the total amount of data available before cuts. $N_{\text{samples}}$ is the number of accepted samples in the \texttt{Mantis Shrimp} dataset. We include notes on the quality cuts applied to each survey, using the specific quality keywords defined in each survey.}
  \centering
 \label{tab:data_table}
  \resizebox{\textwidth}{!}{\begin{tabular}{lllll}
    \toprule
    Source & Ref & $N_{\text{avail}}$ & $N_{\text{samples}}$  & Notes\\
 \hline
     SDSS & \citep{SDSS_four} & 5.1e6 & 2.6e6 & ZWARNING=0,  Class=Galaxy\\
     DESI EDR & \citep{DESI_EDAspectroscopy} & 2.4e6 & 1.1e6 & ZWARN=0, SPECTYPE=GALAXY\\
     DEEP2 & \citep{DEEP2} & 5.3e4 & 3.4e4  & q\_z>2,Cl=G,\\
     GAMA & \citep{GAMA_DR2} & 3.4e5 & 2.0e5 & NQ>=3\\
     VVDS & \citep{VVDS_final_release} & 4.1e4 & 1.0e4  & zflags=3 or zflags=4\\
     VIPERS & \citep{VIPERS_final} & 6.0e4 & 4.9e4 & classflag=0 or 1, zflg<5 \& zflg>=2\\
     6dF & \citep{6dFGS_final} & 1.2e5  & 5.6e4 & QUALITY=3 or QUALITY=4\\
     WiggleZ & \citep{WiggleZ_final} & 2.2e5 & 1.4e5 & Q>=4\\
    \bottomrule
  \end{tabular}}
\end{table*}

Spectroscopic samples were compiled from available surveys and combined together, including the Sloan Digital Sky Survey (SDSS) \citep{SDSS_four} at data release 17, including that of the Main Galaxy Survey (MGS) \citep{SDSS_MGS}, the Baryonic Oscillation Spectroscopic Survey (BOSS) \citep{BOSS_target_guidelines}, and the extended Baryonic Oscillation Spectroscopic Survey (eBOSS) \citep{EBOSS}. We include the Dark Energy Spectroscopic Instrument survey DESI \citep{DESI_EDAspectroscopy} where at time of our compilation, only the early data release was publicly available. SDSS and DESI make up the vast majority of our sample. We also include measurements from DEEP2 \citep{DEEP2}, GAMA \citep{GAMA_DR2}, VVDS \citep{VVDS_final_release}, VIPERS \citep{VIPERS_final}, 6dF \citep{6dFGS_final}, and the WiggleZ \citep{WiggleZ_final} surveys. Information on the quality cuts used for each survey are provided in Table \ref{tab:data_table}, including the final total amount of labels taken from each survey. Our quality cuts and compilation of survey data closely match that of B22 (described in \citep{Beck21PS1STRM}); our main advantage is the inclusion of the sizable DESI EDR spectroscopic sample, adding some 1 million additional targets for training. The targeting guidelines of each of the spectroscopic surveys essentially create a biased and incomplete sample of the input feature space with respect to the entire population of photometrically observable galaxies \citep{NewmanSpectroscopicBias}. We do not consider here how to correct for or flag examples that fall outside the support of the training dataset. We do provide a brief introduction to each survey used in this study in appendix~\ref{appendix:spectroscopic}, including general information about survey goals and what populations they specifically targeted. Spectroscopic catalogs report the target coordinates alongside redshift, so we simply query cutout services at the reported coordinates to build the \texttt{Mantis Shrimp} dataset.

\subsection{Image Processing}
\label{sec:image_processing_mainbody}

\begin{table*}[!ht]
\caption{\small\textbf{Photometric Band Table} Summarizes the properties of the photometric sources used in this paper. Depths are listed as astronomical magnitudes, but the number of objects detected in any particular survey depends on the intrinsic population luminosity of objects in each band. E.g., by visual inspection, many GALEX pointings do not show any signal from the target galaxy.}
  \centering
 \label{tab:photo_filter_table}
  \resizebox{\textwidth}{!}{\begin{tabular}{lllll}
    \toprule
    Source & Ref & Filters & Depth & Pixel Scale\\
 \hline
     GALEX & \citep{GalexSurveyPaper} & FUV, NUV & 19.9, 20.8 & 1.5" \\
     PanSTARRS & \citep{PSSurveyPaper} & g, r, i, z, y & 23.3, 23.2, 23.1, 22.3, 21.3 & 0.25" \\
     UnWISE & \citep{UnWISE_og} & W1, W2 & 20.72, 19.97 & 2.75" \\
    \bottomrule
  \end{tabular}}
\end{table*}

We combine photometric cutout data collected from GALEX \citep{GalexSurveyPaper}, PanSTARRS \citep{PSSurveyPaper}, and WISE \citep{WISE_og} observations (the latter re-processed by UnWISE \citep{UnWISE_og} for filters $W1$, $W2$) for a total 9 different bands of photometry. The Photometric bands are summarized in Table \ref{tab:photo_filter_table}. We chose the PanSTARRS survey for its depth, large footprint, and to ensure our model would be compatible with ongoing surveys using the PanSTARRS telescope \citep{YSE_survey_operations} and related host-galaxy identification software \citep{GHOST}, then chose GALEX and UnWISE surveys for their nearly all sky completion at complimentary wavelengths. \added{Future work would investigate switching PanSTARRS for deeper photometry, e.g., from DESI Legacy Survey or LSST \cite{LSST_main}, when available.}

We are notably missing any near-IR wavelengths, (e.g., that could be provided by 2MASS \citep{2MASS}, UKIDSS \citep{UKIDSS}, or VISTA \citep{VISTA_main}); specifically, 2MASS data is shallow compared to the depth of our targets, \added{while UKIDSS and VISTA are not all-sky surveys so we believed that attempting to include these surveys would increase model complexity by forcing us to appropriately mask our model to these channels when not available. Future work could explore modifying tokenizers for patch-based vision transformers to mask out surveys which are unavailable at any given pointing (as in the AION model, Parker et al. 2025, in prep).}

\subsection{Image Pipeline}
Our dataset is randomly split into 16 nearly-equal sized chunks, and further split into static train-validation-test arrays comprising 70\%-10\%-20\% of each chunk. We then use the \texttt{FFCV} library to compile each chunk and speed up the rate of data loading \citep{FFCV}. During training we use data-augmentation to increase the effective number of samples \citep{AlexNet,DataAugmentation_Review}. We use random flips and rotations of the images then crop around the centers so that each spans a constant 30 arcsec width and height. We had queried each cutout API so that this crop removes any undefined pixels from the rotation. We also randomly sample from our training data with replacement, re-weighting samples from each class to ensure that each class is sampled with the same probability.

\subsection{Data Availability}
The entire dataset, accompanying documentation and tutorials are made available online via PNNL DataHub, with URL found at the beginning of this article. The dataset includes our image vectors as \texttt{numpy} binary FP32 arrays, spectroscopy tables and indices as comma separated value (.csv) files, and other necessary index and mask metadata to make use of the dataset. We also include our compiled \texttt{FFCV} datasets which automatically perform all pre-formatting for the user, and can be used in combination with the PyTorch dataset software created for this project.

\section{Neural Networks}
\label{sec:models}

\subsection{Model Architecture}
\label{sec:early_fusion_models}
\label{sec:model_architecture}
\begin{figure*}[!ht]
    \centering
    \includegraphics[scale=0.6]{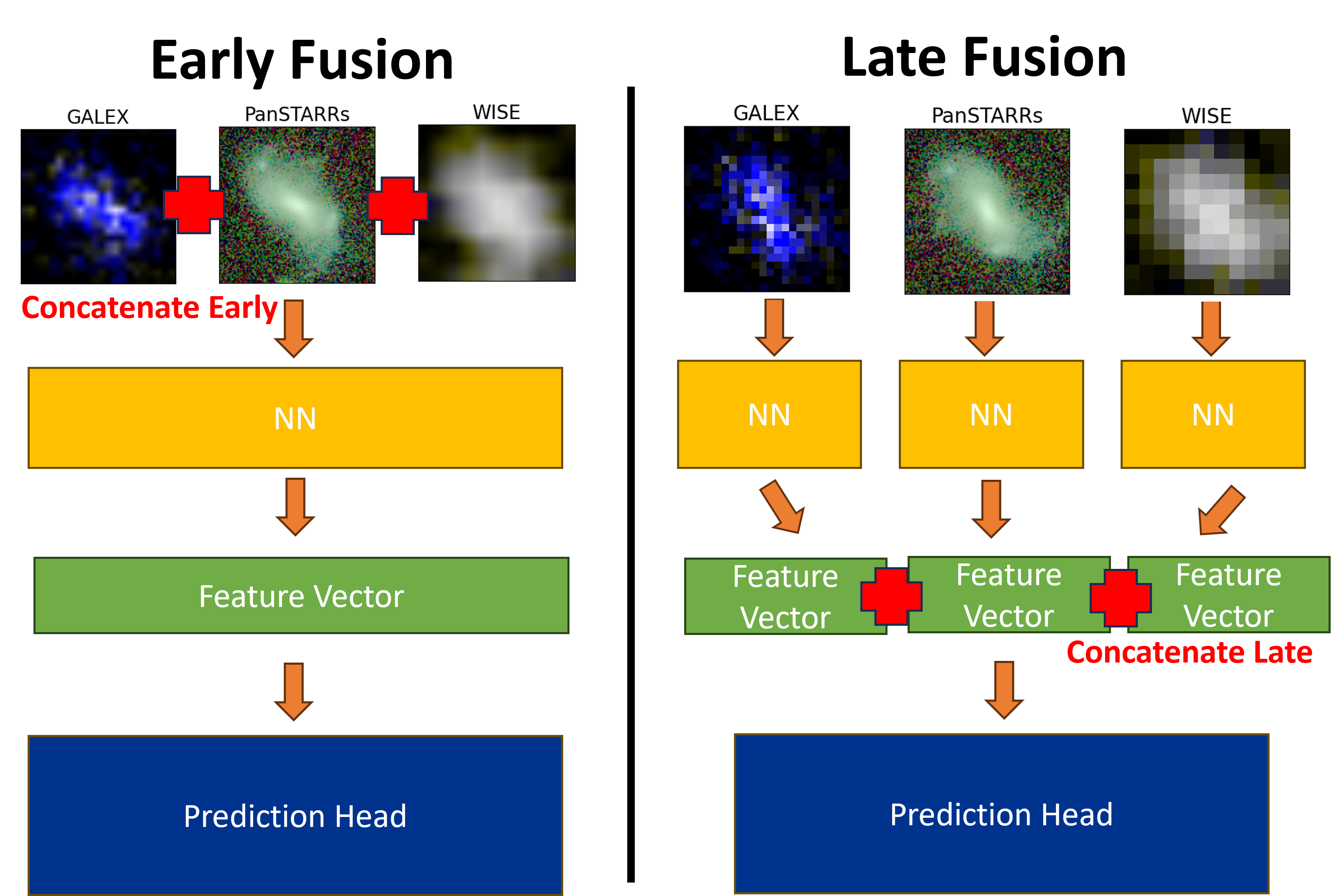}
    \caption{\small \textbf{A visualization of early vs late fusion architectures.} Early fusion (left) upsamples the cutouts to a common pixel grid, then concatenates across channels. Late fusion keeps each image in its original pixel scale, sends the cutouts from each survey through unique feature encoders, and finally concatenates these final feature vectors together to make a prediction.}
    \label{fig:Latevsearly}
\end{figure*}

This study evaluates the effectiveness of early and late fusion models; in this section we will describe both architectures and any specific modifications to the image pipeline. Recall that in early fusion, images are up-sampled to a common pixel grid and then stacked on top of each other, while in late-fusion the cutout from each survey is kept separate and fed through a unique embedding model. The resulting embedding vector is then concatenated together and fed to a densely connected network that maps onto the redshift space. See Figure~\ref{fig:Latevsearly} for a visual depiction comparing the two architectures.

Our early fusion model is a ConvNext-Large \citep{ConvNext} architecture with the input layer modified to accommodate the number of bands in our hyperspectral images. We initialize our weights using the \texttt{ZooBotV2} embedding model weights \citep{ZoobotV2}. To adapt the weights to our architecture, we take an average of the original first convolutional layer across the channel dimension, divide this average by 9, and then replace each of the channel-weights with this reduced average. This choice is reasonable given the convolutional operation performs a sum across channels, so we should approximately retain the same embedding statistics; however, we do not particularly expect the choice to be of great importance as we perform end-to-end finetuning.

For our late fusion model, we use a ConvNext-Small encoder for the GALEX and UnWISE images combined with a ConvNext-Large encoder for the PanSTARRS images. Because there is less weight sharing compared to the early-fusion model, the late-fusion models has a higher trainable parameter count. Our ConvNext-Large encoder is initialized using the same grey-scale weights as the early-fusion model, with the same scheme for modifying the first layer to accept 5-band hyperspectral images. 


\subsection{Accounting for Foreground Dust} Interstellar dust along the line of sight between Earth and the target galaxy preferentially scatters blue-light, which has the effect of reddening the observed color of the galaxy \citep{DustExtinctionReview}. This effect is called dust extinction and can be confused with the reddening due to redshift. Following P19, the dust extinction is incorporated into the neural network following the image-pipeline through feature vector concatenation. Following B22, we concatenate both the extinction from the corrected SFD map \citep{CSFDMap,SFDMap} and the extinction measured from the Planck 2016 map \citep{Planck16_dustmap}, both provided by the \texttt{dustmaps}
library \citep{DUSTMAPs}.

\subsection{Hyperparameter Tuning}

Training neural networks is a process by which model parameters are successively updated to minimize an objective function. ML practitioners typically delineate these learnable model parameters from "hyperparameters" that are set as part of the training algorithm and not updated through the optimization process, e.g., the optimizer's step size \citep{HPO_review}. The value of these hyperparameters have a large influence on the outcome of training, so careful tuning of these hyperparameters must be performed to search for optimal values \citep{HPO_review2}.

We perform hyperparameter tuning by searching over learning rate, hidden dimension size of the classification head, the output dimension (which sets the bin-width of the vector that spans redshift space), and the weight decay parameter, using a simple grid search. We separately run hyperparameter tuning for both our late fusion and early fusion pipeline. Models are evaluated using the performance on the validation set as measured by the point prediction metrics introduced in section~\ref{sec:metrics}. We then train models using the best hyperparameters from both the early and late fusion runs of the tuner. Table~\ref{tab:hparam_tuning} show the values we search over. All hyperparameter tuning runs are allowed a maximum of 60 training epochs. 

\begin{table*}[!ht]
\caption{\small\textbf{Hyperparameter Tuning:} Neural networks notoriously require multiple (often expensive) training runs to evaluate the effectiveness of hyperparameters. We perform a small grid search over manageable hyperparameters to search for better performing models. Future work should investigate platforms to manage this hyperparameter tuning and employ early stopping to speed this process.}
\centering
\label{tab:hparam_tuning}
\begin{tabular}{cccc}
\toprule
    Hyperparameter & Grid Values & Best Early & Best Late \\
    \hline
        Learning Rate &  $[$5e-4$,$3e-4$,$1e-4$]$ & 3e-3 & 1e-3 \\
        Hidden Width & $[2048,1600,1024]$ & 2048 & 2048 \\
        N Classes &  $[500,400]$ & 400 & 400 \\
        Weight Decay &  $[$1e-3$,$1e-4$,$1e-5$,$1e-6$,$1e-7$]$ & 1e-7 & 1e-7 \\
     \hline
    \bottomrule
  \end{tabular}
\end{table*}

\subsection{Training}

Using the best hyperparameters selected from both our early and late fusion hyperparameter tuning runs (shown in Table~\ref{tab:hparam_tuning}), we train a fiduciary model combining each of our photometric surveys. We use the NAdam optimizer~\citep{NAdam} to minimize cross entropy loss. We train our models for 150 epochs, stopping after an estimated 2 days of total training time on 16 NVIDIA A100 GPUs (40 GB VRAM) distributed across 2 compute nodes. We anneal the learning rate throughout training using a learning rate scheduler tracking the validation cross entropy loss, which reduces the effective learning rate as $\text{LR}_{\text{new}} = \text{LR}_0 \times 0.5$ if the validation loss is not seen to improve over a period of six epochs. We use a batch size of 16. We sample from our training dataset using a weighted random sampler with weights set to provide a uniform sampling over the class occurrences. Because we sample from the training data with replacement, its possible to see the same sample more than once per epoch. We set a static 1000 total batches per epoch. We track both the loss and our evaluation metrics at regular intervals on the validation set.

After training fiducial models that combine all the photometric bands together, we train three more sets of models where we ablate the IR, the UV, or both the IR and UV in order to better understand if/how our model utilizes the information from each band. This ablation is performed by simply removing the bands from the data input in the case of the early fusion model. For the late fusion model both the data and the encoders used to learn latent feature vectors are removed-- in this case, we note that the total number of model parameters does change with this ablation, which creates a confounding variable. This is unfortunately unavoidable.

\section{Interpretability}
\label{sec:interpretability_methods} 
In this section, we share our specific implementation of Shapley values. Much of the discussion centers on the choice of baseline image $b'_i$, or how different choices of $x'_i$ can be used to help probe different model behavior.

\subsection{Shapley Values:}
\label{sec:shapley_methods}

In this work, we apply Shapley values to understand the contribution from individual photometric bands. For each datapoint, we can modify each band individually to suppress information about the target galaxy from that band, then measure the resultant change in the predicted redshift from our ablation. The exact choice of how to modify the input band is critical and affects the interpretation of the result \citep{DistilFeatureBaselines}. Luckily, by applying our domain knowledge, we will introduce a well-motivated choice for the creation of ablated baseline images, $b'_i$, that can be easily constructed using familiar astronomical python libraries.

We modify the input band by using the \texttt{sep} python library \citep{seplibrary_Barbary2016}, a python-binding of \texttt{Source Extractor} \citep{sextractor}, to identify the sources in an image with a $2\sigma$ detection limit. Because we know that the cutouts were centered on the astrometry of the target galaxy, we take the object whose position is nearest to the center of the cutout as the target galaxy. If no galaxy is detected within a 5 pixel length to the center, we instead use a static aperture which we describe in the next paragraph. The \texttt{sep} library returns best-fit ellipses to the objects: using the ellipse of the target galaxy, we mask out the target galaxy by scaling the ellipse to 4x the Kron radius measured in each band \citep{KronRadius}. We then fill in this mask with \added{values sampled from a model of the sky. We model the sky using a Gaussian and estimate that Gaussian's moments using robust statistics, using the median and normalized median absolute deviation rather than the mean and standard deviation respectively. We compute these moments only on pixels not containing a source as identified by source extractor.} We allow the resulting image to represent the ablation of features of the target galaxy in that band. When computing Shapley values, we modify our image by replacing the original band images with the feature ablated band image. This choice keeps the pixel values of other sources and the background noise static. Because Shapley values are coupled to the \textit{difference} between the NN output on original and feature-ablated images (see Eq.~\eqref{eq:shapleyvalue}), this ensures only pixels related to our target galaxy influence the Shapley value.

As mentioned in the preceding paragraph, we use a threshold on the maximum allowed pixel distance of a source from the center of the cutout to be considered a detection of the target galaxy. If no sources are detected satisfying this constraint, we instead use a small circular aperture of pixel radius 2,4,2 for GALEX, PanSTARRS, and WISE, respectively. This small aperture choice is made to eliminate any possible localized central luminosity that exists below our 2$\sigma$ detection threshold. We then proceed to mask out this central region and fill with Gaussian noise sampled from the image. 

\begin{figure*}[!ht]
    \centering
    \includegraphics[scale=0.55]{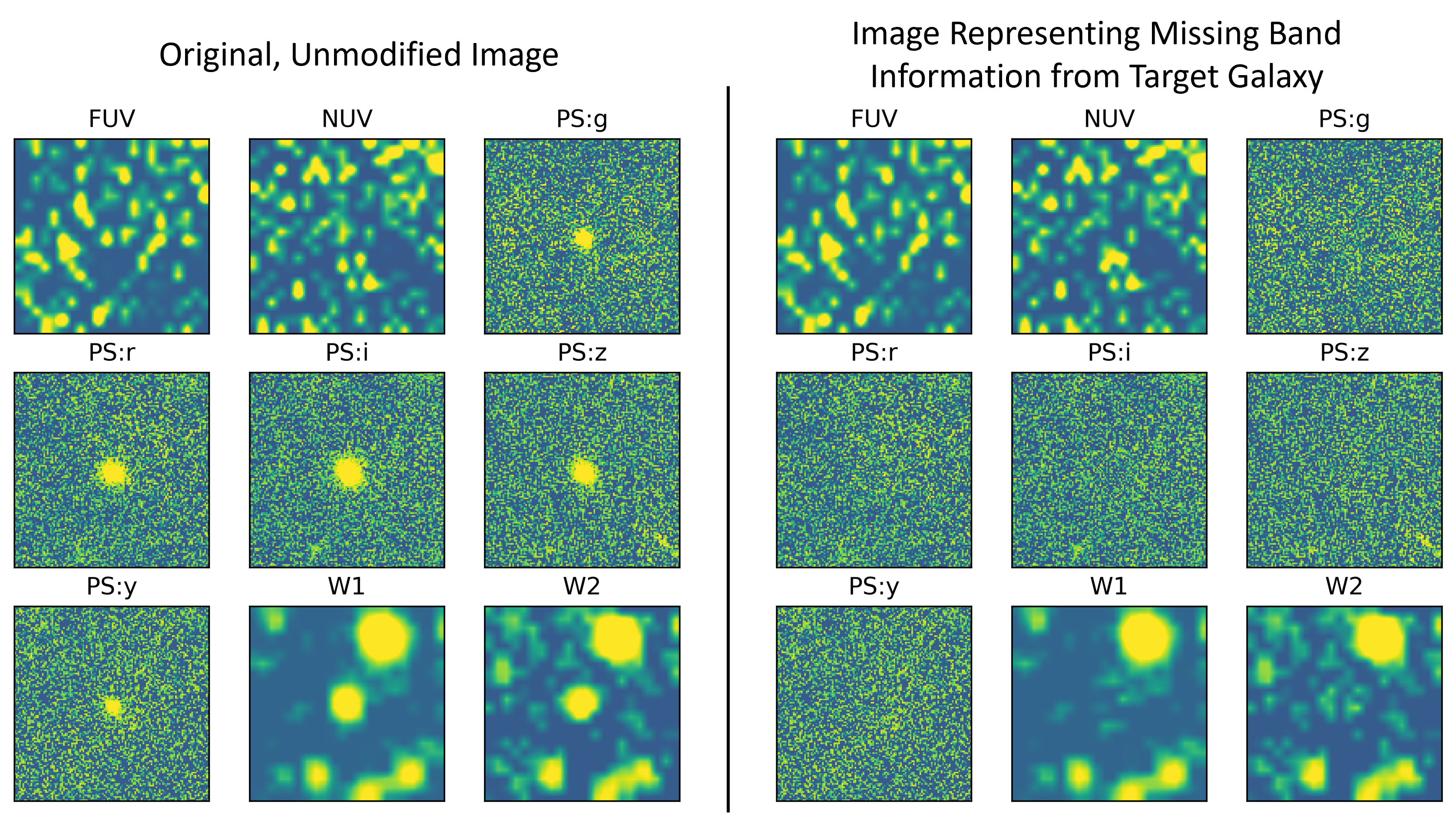}
    \caption{\small \textbf{Example of using source extractor to define baseline image}. On the left we provide an example of the unmodified original image in each instrument band. The source is centered as a result of our cutout pipeline. Because we know this to be true across every cutout, we can use source extractor to mask out the center-most source and fill with Gaussian error sampled from the image, which is shown on the right. The per-channel result of masking out the center most source is used to to define a null contribution of features from each filter.}
    \label{fig:sep}
\end{figure*}

In Figure~\ref{fig:sep}, we show an original unmodified image of a target galaxy across each of the 9 bands, and the modified version where we have removed the galaxy using \texttt{sep}. A few additional examples are provided in appendix~\ref{appendix:sep_examples}. We leave a rigorous evaluation of this process to future work, but believe because we have focused the Shapley values onto the target galaxy pixels, this is a more reasonable starting point for Shapley Values compared to naive options like simply replacing the entire filter with zeros. We will discuss in the limitations and future work section what improvements could be made to the pipeline. \added{In addition, we include appendix~\ref{appendix:ShapleyBlends} that evaluates the robustness of our approach to blends.}

\section{Metrics}
\subsection{Evaluation Metrics:}
\label{sec:metrics}
Recall that our model produces a probability density of redshift for each input, and that we define a point-estimate from these probability densities as the expectation value. Following standards in the field\footnote{See the metrics defined for the \href{https://rail-hub.readthedocs.io/projects/rail-notebooks/en/latest/rendered/evaluation\_examples/Evaluation\_by\_type.html}{LSST photometric redshift evaluation}}
we evaluate our model using both point-like metrics and evaluate the probability densities produced by our model via cumulative density calibration metrics. For the point based metrics, we define the following using the set of scaled residuals $r = {\frac{z_i - \hat{z}_i}{(1+z_i)}}$, and allow $r_i$ to be the $i$-th element of that set. In the following let $\text{med}(\cdot)$ represent the median over a set. We define a robust measure of spread, the scaled median absolute deviation (\textbf{NMAD}) $= 1.4826 \times \text{med}(|r_i - \text{med}(r_i)|)$, the bias of the residuals (\textbf{BIAS}) $= \frac{1}{|r|} \times \sum_{i=1}^{|r|} r_i$, and the percentage of catastrophic outliers ($\boldsymbol{\eta}$) defined to be the percentage of scaled residuals with values greater than 0.05. \added{We note that the exact value at which a residual is considered a catastrophic outlier varies from work to work, but the most common value is 0.15. To ease comparison with future works, we also include a calculation of catastrophic outliers greater than 0.15, $\boldsymbol{\eta_{>0.15}}$, and report it in appendix~\ref{appendix:eta015}.}

We report uncertainties in these point estimate statistics by resampling the set of residuals using the bootstrap strategy with 100,000 resamples. Other sources of uncertainty are not factored into the values we report: one important factor often overlooked would be the variation induced by randomness in the gradient descent training process and model initializations. Understanding the effects of early and late fusion rigorously would require drawing samples from the distribution of models that could be trained from either choice, i.e., training multiple models from scratch with different seeds \citep{ModelVariabilityOverTraining}. However, sampling from the distribution of models this way is very expensive on the scale of this work and beyond the scope of this pathfinder study.

\label{sec:PIT}
Conditional density estimation and proper calibration is an active area of study in photometric redshift estimation. Generally, neural networks do not accurately quantify uncertainty \citep{ML_Uncertainty_Review}. Prior works in the photometric redshift field \citep{PZEvaluationLSSTSchmidt2020} have evaluated calibration using the probability integral transform (\textbf{PIT})\citep{PITOGDawid1984,PITAstroPolsterer2016}, and a ``CDE loss'' \citep{PZEvaluationLSSTSchmidt2020}; we will introduce both in turn in the following paragraphs.

The PIT is a histogram of the occurrences of true redshift in the set of CDFs measured from our conditional density estimates. While in truth our model estimates discrete probability mass functions, in the following we will allow these to approximate the probability density function $F(x_i,\;\theta) ~ \hat{p}_i(z)$, so that we can define the cumulative density as $\text{CDF}(z_i,\hat{p}_i) = \int_0^{z_i} \hat{p}_i(z') \; \mathrm{d}z'$. If $\hat{p}_i(z)$ equals the true PDF, $p_i(z)$, then the histogram visualizing the density of the set of CDFs evaluated at the true redshift, $\{\text{CDF}(z_0,\hat{p}_0),
\ldots,\text{CDF}(z_N,\hat{p}_N)\}$ will be flat. It is important to remember that PIT is evaluated on the entire ensemble of our test dataset and does not guarantee the probabilistic calibration of any individual estimation.

As a single scalar metric of probabilistic calibration we will also calculate the conditional density estimate loss (\textbf{CDE loss}) test statistic \citep{CDElossIzbicki2016}. The CDE loss is an approximation to the squared error between the true conditional density of redshift and the estimate our model produces, up to a constant that depends on the true conditional density. Both the PIT visual metric and the CDE loss are included in the LSST RAIL standard suite of benchmarks \citep{LSST_RAIL_Photoz}. 

Finally, we also re-calibrate our final model using the \texttt{CalPIT}
library \citep{Dey_ConditionalCalibration2022_CalPIT}. This re-calibration technique trains an auxiliary network to learn a local (in feature space) calibration of the initial CDE estimate to improve overall performance on our total CDE calibration. So as not to confound the effects of our ablation study, we use this re-calibration only on our final model that we expose to users in the web-app. Users of our web-app may optionally apply the calibration from \texttt{CalPIT}. We provide additional detail on \texttt{CalPIT} and re-evaluation in appendix~\ref{appendix:CalPIT}.

\section{Results}
\label{sec:results}
\subsection{Early vs Late Fusion:}
\label{sec:result_ablation}
To compare the effectiveness of early vs late fusion architectures, we perform two independent hyperparameter searches then train both an early fusion and late fusion model at the best set of values found during the hyperparameter search. We then ablate the different surveys and re-train, which allows us to directly evaluate whether adding additional surveys together will increase the overall performance of the model. We compare models in Table~\ref{tab:ablation_results} using the suite of metrics defined in Section~\ref{sec:metrics}, which includes the scaled median absolute deviation (NMAD), the bias of the residuals, and the percentage of catastrophic outliers, $\eta$, and the conditional density estimate (CDE) loss. 

The first key takeaway from our results is that the early and late fusion architectures show very similar performance improvements in each group of ablation experiment. In fact we will show in the following sections that overall the early and late fusion models perform and behave very similarly. The inclusion of IR data makes a more significant improvement over the optical only model than the inclusion of the UV data; however, we also show that including both UV + Optical + IR gives marginally better results than only IR + Optical. Most importantly, these performance improvements indicate that each fusion strategy is successfully integrating data from each survey. The fact that the UV data is not as useful as the IR data should be expected given the depth of GALEX and relative low surface-brightness of the majority of our sample in the UV. \added{It should also be expected given it is known that $r - W1$ color is indicative of redshift \cite{DESI_LRG}.} The UV data nonetheless 1) forms an excellent contrast to the inclusion of the IR and 2) does seem to add a marginal improvement so is included in our final model. We note that an earlier version of this work, \citep{Engel2024PreliminaryRO}, had performed the same ablation study (for early fusion only) using a 100$\times$ smaller model (by learnable parameter count) and had not observed that the inclusion of UV improved upon Optical+IR models, suggesting that parameter count may be an important factor for determining whether the model learns to ignore low-frequency features in the UV or utilize them.

\begin{table*}[!ht]
\caption{\small\textbf{Ablation Model Performance:} We evaluate our model on the test dataset, comparing both early and late fusion models and ablating UV and IR filters, using the metrics described in Sec~\ref{sec:metrics}. Overall, each early and late architecture performs very similarly to their counterpart given the same data inputs. The improvement in performance compared to the optical only model shows that our models are able to incorporate information across each of the photometric bands. We confirm that the best models across all point-wise metrics are those trained with all the available photometric bands, followed closely by those trained with Optical and IR. This is understandable given that we expect only the most nearby objects to have detections in the UV.  \textbf{Values reported with leading digit of uncertainty as parenthetical}}
\centering
\label{tab:ablation_results}
\begin{tabular}{llcccc}
\toprule
Fusion & Data Included & NMAD ($\times 10^{-2}$) & Bias ($\times 10^{-2}$) & $\eta$ (\%) & CDE loss \\
\hline
Neither & Optical only & 3.507(5) & 2.07(1) & 25.33(5) & -4.53(3) \\
Late & Optical and UV & 3.307(5) & 1.69(1) & 22.90(5) & -6.00(4) \\
Early & Optical and UV & 3.316(5) & 1.37(1) & 22.91(4) & -3.90(5) \\
Late & Optical and IR & 2.527(4) & 1.81(1) & 19.20(4) & -8.84(4) \\
Early & Optical and IR & 2.531(4) & 2.10(1) & 19.02(4) & \textbf{-9.68(2)} \\
Late & Optical, UV, and IR & 2.461(4) & 1.46(1) & 17.79(4) & -8.81(4) \\
Early & Optical, UV, and IR & \textbf{2.438(4)} & \textbf{1.151(9)} & \textbf{17.60(4)} & -7.45(7) \\
     \hline
    \bottomrule
  \end{tabular}
\end{table*}

\subsection{Performance on Relevant Sub-Populations}
\label{sec:results_subpop}
In addition to comparing early vs late time fusion variants of our model, we compare the \texttt{Mantis Shrimp} model against several literature benchmarks on sub-populations within our test dataset. The results are available in Table~\ref{tab:ablation_results}. We emphasize, our model is \textbf{not} re-trained or fine-tuned on these populations, but rather, we mask down our original test dataset to these sub-populations. There are four relevant sub-populations which we next describe. 

\subsubsection{SDSS Main Galaxy Sample}

 The first is the SDSS main galaxy sample (MGS). The SDSS MGS is a flux limited sample taken from the SDSS spectroscopic survey with limits on \texttt{PetrosianMag\_r} $< 17.77$ \citep{SDSS_MGS}. Due to this, the SDSS MGS is a much more nearby sub-population compared to our total test dataset with redshifts typically less than 0.4. Since P19 and the subsequent publication of their dataset \citep{Dey2021Capsule}, multiple works have shown that CNNs provide significant performance boost over tabular magnitude/color models, (one such example given as the official SDSS redshift model \citep{Beck2016LLNSDSS}). Due to multiple works having published performance and for the fact that the dataset is publicly available, SDSS MGS is an interesting benchmark for comparison. The SDSS MGS sub-population makes up 12\% of our original test dataset. We re-emphasize that for the SDSS MGS comparisons, all other works utilize the SDSS optical bands as input to their model, while we continue to use the GALEX + PanSTARRS + UnWISE photometry (Notably, we lack an analogue for the SDSS:u band, and they lack the WISE:W1 and WISE:W2 bands). In addition, those marked with an asterisk(*) in Table \ref{tab:subpop_reults} are reported directly from the authors on a similar population of galaxies, but are not verified in this work nor are an exact match onto the sub-population present in our test dataset. Despite these caveats, we believe that their performance benchmarks are valid and relevant for comparison; for example, for a user deciding which tool to utilize in their work.  

While we replicate the observation that computer vision models are more performant than tabular based models (represented here by \citep{Beck2016LLNSDSS}), we are unable to match or exceed the performance quoted by other computer vision works. We speculate that this could be due to 1) the specificity of training that is performed in each work on the SDSS MGS (whereas in our work we mask down our test sample to match onto this sub-population). 2) Prior works train an ensemble of models and report results that are an average over the ensemble's output, whereas this work trains a single model only. 3) The differences in the photometric bands being utilized across works: it is typical to use SDSS data to train on the SDSS MGS subpopulation and so would utilize the SDSS:u band. Nonetheless, we do meet a performance of NMAD<1e-2 with our early computer vision model, which was thought in the community to be an impassible barrier for performance on the SDSS MGS and an important motivating moment for the computer vision models used in this work, see P19. 

\subsubsection{PS1 X WISE}
A second sub-population is that of the observed galaxies after performing a join between PS1 and WISE, which we call PS1 $\times$ WISE sub-population. We approximate this sub-population by using the galaxies at which B22 provides a redshift, (and so implicitly, we make use of that work's star-galaxy classifier). This sub-population makes up just 69\% of the original test dataset, and is interesting insofar as B22 is the most relevant comparison of a tabular based photometric redshift model utilizing both PanSTARRS and WISE. We emphasize here that we make no effort to restrict our matched sample to galaxies that B22 would have trained upon. In addition, because we are matching on all galaxies that B22 provides an estimate some amount of the sample will be on the population of the DESI EDR galaxies. This represents a population shift from what B22 would have been trained upon. Additional details regarding how we separate these sub-populations from the global test dataset are available in appendix~\ref{appendix:comparison_details}. 

Overall, we achieve comparable results to B22 on the PS1 $\times$ WISE sub-population. The most dramatic difference in behavior is on the CDE loss measuring probability calibration of our conditional density estimates. We will elaborate in the next section upon this behavioral difference, and elaborate on potential use case differences between our two models in the discussion. 

\subsubsection{PS1 Objects}
The third sub-population comes from matching onto the population described in \citet{tarrio}, which provides a catalog of photometric redshifts for all PanSTARRS objects observed in the first data release, (unlike B22, without any cuts on star/galaxy separation or feature space) and included a null value in training to handle objects with missing photometry in some bands. We form the sub-population by performing a cross-match of our test dataset onto their catalog with a 1.5" tolerance. This cross match retains 77\% of our sample. \textbf{Notably, this implies that roughly 23\% of our sample is not actually detectable in the PanSTARRS survey!}

We observe significant improvements in NMAD, $\eta$, and CDE Loss on this sub-population compared to the earlier work of \citet{tarrio}, but given our inclusion of infrared photometry and the addition of DESI EDR in our sample, we are effectively forcing the \citet{tarrio} model to extrapolate beyond its original training population, which is expected to be disastrous for their performance. Due to this effect, one could argue that a more fair comparison would be to restrict samples to a shared feature space (for example, using magnitude cuts); however, without a motivating problem to define a relevant feature space, we leave it to authors who know their problem to make their comparison.

\subsubsection{DESI Legacy Survey DR9}
The final sub-population is created by matching onto the DESI Legacy Survey DR9 Photometric redshift catalog \citep{Zhou_Photoz_DECaLS_DR9}. This catalog uses a tree-based algorithm on colors from the DESI Legacy Survey's visual and UnWISE bands, extinction, and a morphological parameter, for a total of just 9 features. This amounts to a radical decrease in complexity compared to our convolutional neural network. In addition, \citet{Zhou_Photoz_DECaLS_DR9} uses nearly the same spectroscopic surveys as this work, including the DESI EDR spectroscopy. To create our matched set, we query the DESI Legacy Survey \texttt{ls\_dr9.photo\_z\_9p1} table for all galaxies that have a spectroscopic redshift value. We then remove the galaxies that were in the training set of \citet{Zhou_Photoz_DECaLS_DR9}. We perform a cross-match onto the galaxies that were in our own test-dataset (with a 1.5" tolerance), which results in 281,281 samples. We institute a maximum redshift difference between our own spectroscopic labels and those from the table of 1 part in 1000, which removes an additional 178 samples, for a final 281103 cross matched galaxies. We use the \texttt{ls\_dr9.photo\_z\_9p1.z\_phot\_mean} value as  \citet{Zhou_Photoz_DECaLS_DR9} predicted redshift and their \texttt{ls\_dr9.photo\_z\_9p1.z\_phot\_std} value for their uncertainty, which notably makes the simplifying assumption that the reported photo-z PDFs are Gaussian distributed.

Compared to the DESI Legacy Survey's (hereafter LS) photometric redshift model, our performance is considerably worse on a matched sample. This is especially troubling given the fact that (as mentioned in the preceding paragraph) the LS photometric redshift model is considerably less complex and only uses 9 tabular features. \added{One intuitive explanation is that the LS photometry is about a magnitude deeper in each optical band than the PanSTARRS photometry. We also consider that the DESI spectroscopic survey's selection criteria rely on LS photometry, ensuring that all objects from DESI EDR are ensured to be detected in LS, but not necessarily in PanSTARRS. We expand upon this in the discussion section.}

\begin{table*}[!ht]
\caption{\small\textbf{Model Performance Across Sub-populations:} We compare the \texttt{Mantis Shrimp} model to prior works on various sub-populations from literature. The asterisk (*) marks metrics as reported by the original authors, rather than computed directly. Some previous works do not report a value for the CDE loss, which we indicate with N/R, or ``not reported.'' We report values for catastrophic outlier \textbf{$\eta$} as residuals greater than 0.05; for values of \textbf{$\eta_{>0.15}$} see appendix~\ref{appendix:eta015}.  \textbf{Values reported with leading digit of uncertainty as parenthetical}}
  \centering
 \label{tab:subpop_reults}
  \begin{tabular}{llccccccc}
    \toprule
    Subpop. & Model & NMAD ($\times 10^{-2}$) & Bias ($\times 10^{-3}$) & $\eta$ (\%) & CDE loss \\
    \hline
    \multirow{6}{*}{SDSS MGS} & Multi-instrument CNN, early (This Work) & 0.995(4) & -0.19(4) & 0.53(2) & \textbf{-25.9(1)}  \\
    & Multi-instrument CNN, late (This Work) & 1.023(4) & -0.50(4) & 0.59(2) & -25.0(1) \\
    & Optical-only Tabular \citep{Beck2016LLNSDSS} & 1.354(6) & 0.62(6) & 1.37(3) & -19.61(2)  \\
    & Optical-only CNN \citep{Pasquet2019}* & 0.91 & 0.1 & 0.31 & N/R   \\
    & Optical-only CNN \citep{Hayat_2021_selfsupervised}* & \textbf{0.83} & 0.1 & 0.21 & N/R \\
      & Optical-only CNN \citep{Dey2021Capsule}* & 0.89 & \textbf{0.07} & \textbf{0.19} & N/R   \\
     \hline
     \multirow{3}{*}{PS1 $\times$ WISE}  & Multi-instrument CNN, early (This Work) & 1.862(3) & 2.88(5) & \textbf{6.59(3)} & \textbf{-13.08(5)} \\
     & Multi-instrument CNN, late (This Work) & 1.882(3) & 3.46(5) & 6.74(3) & -12.91(4)  \\
      & Multi-instrument Tabular \citep{WISE-PS1-STRMBeck22} & \textbf{1.830(3)} & \textbf{-0.40(6)} & 9.00(4) & -10.45(1)   \\
     \hline
     \multirow{3}{*}{PS1 Objects}  & Multi-instrument CNN, early (This Work) & \textbf{1.964(3)} & 4.72(5) & \textbf{8.72(3)} & -12.94(3) \\
     & Multi-instrument CNN, late (This Work) & 1.989(3) & 5.64(6) & 9.00(4) & \textbf{-12.96(2)}  \\
      & Optical-only Tabular \citep{tarrio} & 3.527(6) & \textbf{1.63(8)} & 24.00(5) & -6.80(1) \\
     \hline
     \multirow{3}{*}{DESI LS DR9} & Multi-instrument CNN, early (This Work) & 1.960(5) & 5.43(7) & 7.93(5) & -11.96(3) \\
      & Multi-instrument CNN, late (This Work) & 1.977(5) & 6.04(7) & 8.18(5) & -11.86(3)   \\
      & Multi-instrument Tabular \citep{Zhou_Photoz_DECaLS_DR9} & \textbf{1.508(4)} & \textbf{-5(1)} & \textbf{6.09(4)} & \textbf{-14.36(3)} \\
     \hline
    \bottomrule
  \end{tabular}
\end{table*}

\subsection{Probabilistic Evaluation}
\begin{figure*}[!ht]
    \centering
    \includegraphics[scale=0.49]{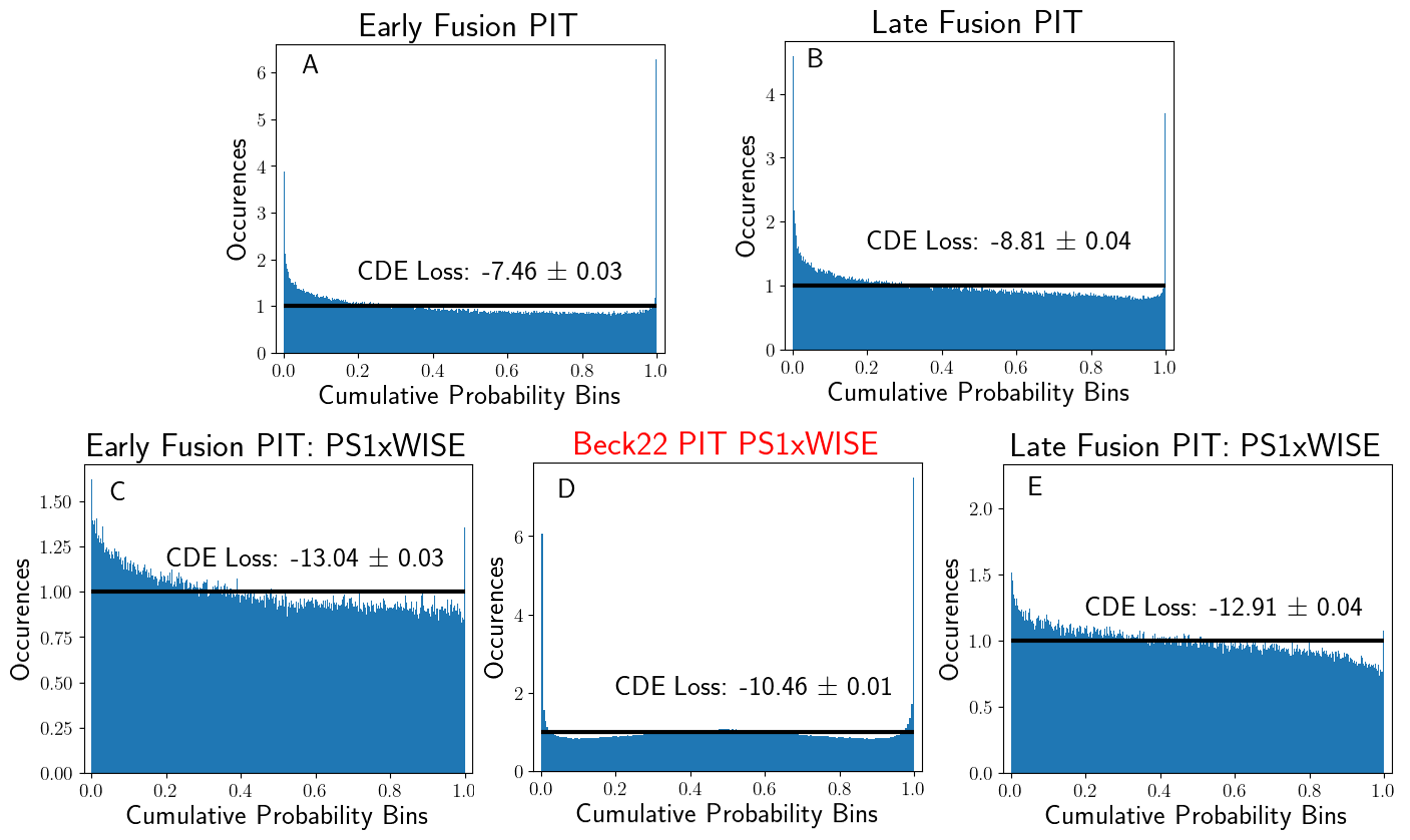}
    \caption{\small\textbf{PIT} The rate of true occurrences in the CDF across the ensemble of test datapoint PDFs generated by our model, where a perfectly calibrated model would have all histogram bins line up across the horizontal black line. In the top row we compare our early and late fusion models. Overall, both models have a higher occurrence at low cumulative probability and near the median than should be expected from a perfectly calibrated model. In the bottom row (Figure C) we compare to the B22 model on the PS1 x WISE galaxies, which is the most comparable work utilizing the PanSTARRS photometry as used in this work. The sharp histogram bins in B22 at 0.0 and 1.0 is indicative of uncertainty that is under-estimated. We emphasize here that the sample of PS1 $\times$ WISE galaxies include some amount of the training sample B22 used and that a significant portion would come from DESI EDR matches that B22 did not train on.}
    \label{fig:PIT}
\end{figure*}

 We evaluate the probabilistic calibration of our models through the visual PIT metric, shown in Figure~\ref{fig:PIT}. The calibration is evaluated across the entire ensemble of the test dataset. We also compare to the calibration of the B22 model through their self-reported uncertainty estimates. We find that overall, late and early fusion models exhibit similar calibration behavior. We also find our models are much better calibrated on the ensemble of our entire test dataset than B22, as evidenced by both the PIT plots and lower CDE loss.

\begin{figure*}[!ht]
    \centering
    \includegraphics[scale=0.55]{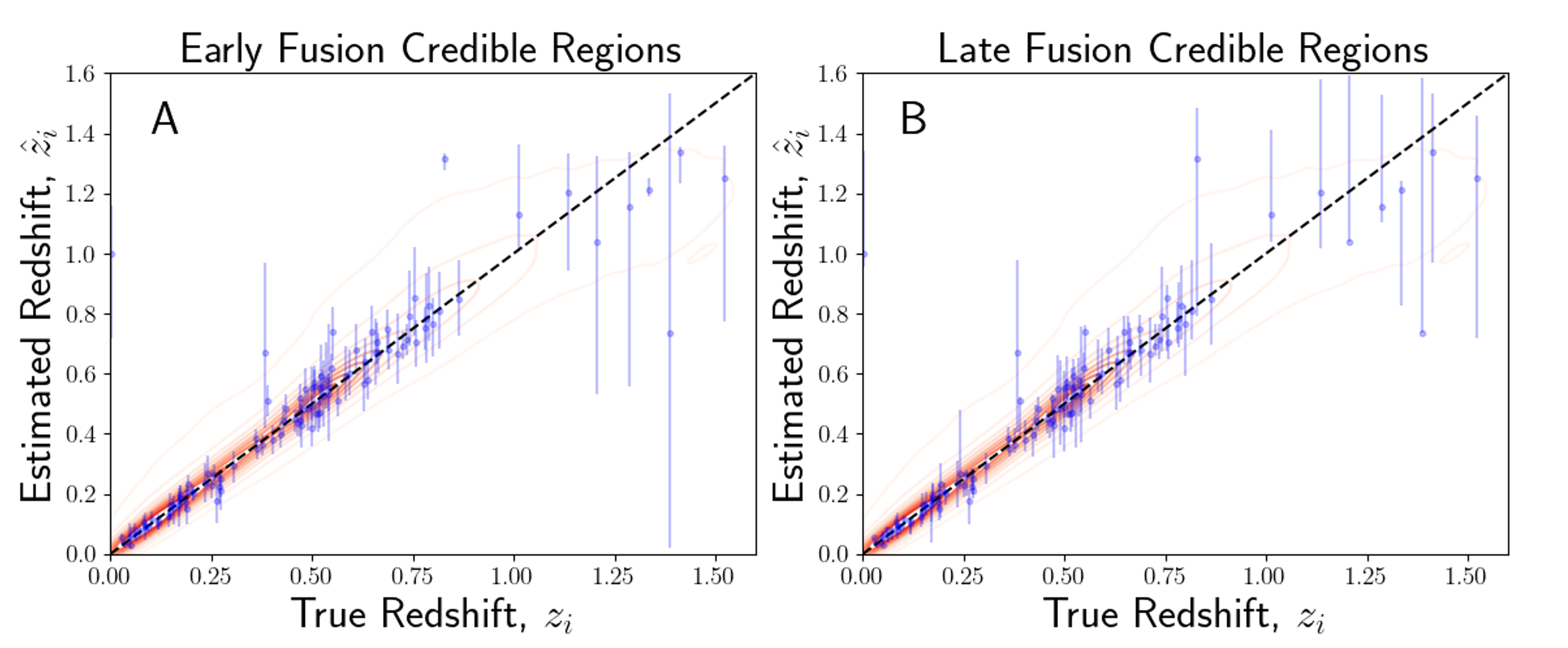}
    \caption{\small\textbf{point-to-point clouds} A: Early Fusion. B: Late Fusion. Both plots show the point-estimates (y-axis) against the ground truth redshift values (x-axis) visualized as a kernel density plot. The Blue points with error bars are a random selection from the population of test points, where the error bar shows a confidence region estimated from our PDFs. The chosen random sample of points is shared between the two plots.}
    \label{fig:p2p}
\end{figure*}

Our probabilistic interpretation of the output of our neural network allows us to create confidence regions unique to each point prediction, as demonstrated in Figure~\ref{fig:p2p}, where we overlay the point estimates to true redshift with a small random sample of the test points with error bars demonstrating the 90\% confidence region extracted from the CDE. We also visualize the overall density of predictions of the model using a kernel density estimate. It should be noted that these results are currently limited to qualitative observations of trends, and we expand upon these limitations and the future of interpretability in the discussion.

\subsection{Shapley Interpretability:}
\label{sec:results_shapley}
\begin{figure*}[!ht]
    \centering
    \includegraphics[scale=0.45]{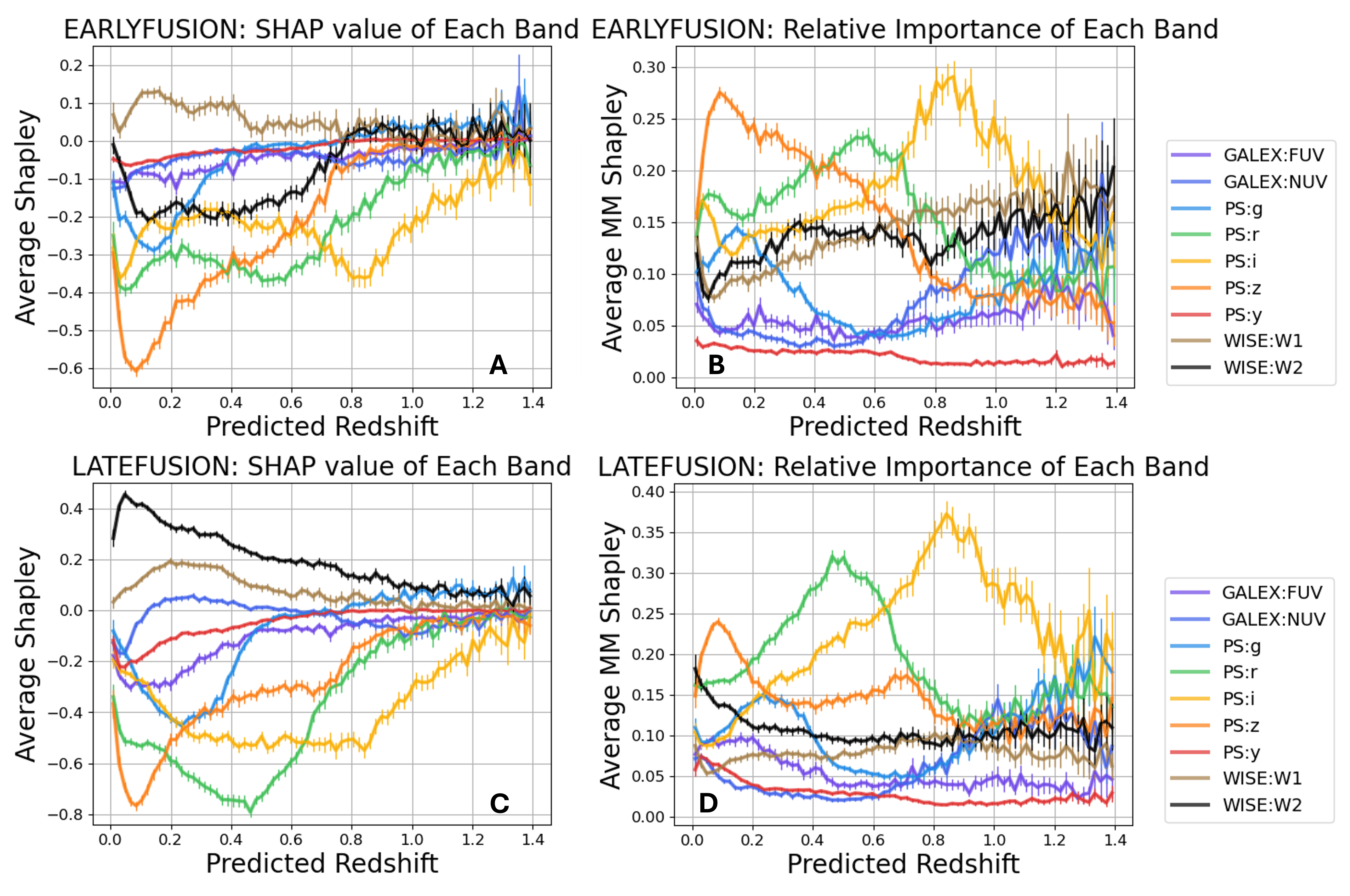} 
    \caption{\small\textbf{Shapley (left column) and MM-SHAP (right column) values for each band for (top row) Early and (bottom row) Late fusion, averaged across the samples in each redshift bin.} A simple intuition for reading these charts is as follows: A positive (negative) Shapley value means that as the flux in the band is increased the model predicts higher (lower) redshift. Guided by this intuition, we make some observations. First: we observe that in the interval $0<z<1.2$ both architecture exhibit highly structured Shapley values. Outside this range at redshift $z>1.2$, the structure disappears and all bands have higher variance. Second: we observe some similar behavior between the two models in the Optical bands, but very different behavior in the UV and IR bands. Most strikingly, the average Shapley value for the W1 band is positive for late fusion and negative for early fusion on a large interval $0<z<0.8$.} 
    \label{fig:interp}
\end{figure*}

We compute both the original Shapley and the normalized MM-SHAP scores for each filter as outlined in section \ref{sec:shapley_methods}, and visualize the results in Figure~\ref{fig:interp}. Recall that Shapley values represent answers to the question "\textit{how is the presence of this band impacting our photo-z estimate?}", while MM-SHAP answers the question "\textit{what is the relative importance of each band?}". Our goal in this section is simply to advocate for the use of interpretability methods from AI literature evaluated from the perspective of our domain expertise.

These Shapley values paint a picture about the model's usage of each band and the way each architecture is sensitive to ablations. To pick a few points to highlight, we see that the Shapley values are highly structured at low redshift and then converge to a high variance and less structured scatter for redshifts z>1.2. Given that Shapley Values are defined specific to each datapoint it should be noted that no average or shared structure across images needed to have emerged; i.e., the error bars being small relative to the differences between bands is indicating a global structure to how the model utilizes each band at each redshift. Critically, we observe shared trends between the early and late fusion architectures in how they utilize the optical and W2 bands. But, we also observe an increased sensitivity to ablations, indicated by higher overall Shapley values, in the late fusion model. This could be explained by the fact there is less weight sharing in that architecture--- because more of the architecture must specialize to a smaller set of bands, ablating those bands can have a larger impact compared to the early fusion architecture.

Both architecture's optical bands are negative and the highest Shapley value band over much of the interval are an IR band. We argue this represents the intuition we presented in section~\ref{sec:physics} on the physics of redshift: as the galactic flux density curve is redshifted generally more flux density becomes concentrated in the IR wavelengths and moves out of the optical wavelengths. If we frame the presence of a band as a brightening of flux in that band (relative to some static optical band), we should expect that a brightening of the IR bands would lead to a higher redshift estimate (and therefore a positive Shapley value) and likewise, a brightening of the optical bands (relative to static IR bands) would lead to a lower redshift estimate and therefore to a negative Shapley value. We also see evidence of the $g$ band importance (best observed in the MM Shapley value plots) falling quickly at redshift of z=0.4, which aligns with when the 4000\AA~break has moved out of the $g$ band.

\section{Discussion}
\label{sec:discussion}
Our research study is framed as a pathfinder to determine the advantages and disadvantages of early and late fusion architectures, as an description of a newly available tool for photometric redshift conditional density estimation, and as an exploration of how tools developed for AI interpretability could be combined with domain knowledge to probe learned behavior of NN models \added{in astronomy.} We discuss each of these mains points in turn.

\subsection{Late vs Early Fusion}
While we do observe statistically significant differences in metrics computed from our late and early fusion architectures, the overall performance is very similar compared to other works. Because of this fact, we do not interpret this to be any impactful advantage in performance or behavior between early and late fusion architectures for the photometric conditional density estimation task. Our ablation experiment confirms that both architectures are successful in combining information from multiple surveys together compared to a baseline optical only model. The main trade offs between early and late fusion architecture are efficiency for re-usability: the early fusion architectures is more weight efficient than the late fusion because weights are shared in the convolutional layers. This leads to an overall smaller training time. Late fusion model's main advantage would be the potential for re-use: if a late-fusion model could be trained that acted as an arbitrary feature encoder for each photometric survey then multiple projects could re-use encoding from a single model depending on the task. See for example general feature encoders for galaxy morphology in \citet{ZoobotV2}. 

\subsection{Specific Use Cases for \texttt{Mantis Shrimp} and Future of CNNs for Photometric Redshift}

Our CNN-based approach to photometric redshifts offers several unique advantages, even as it highlights areas for future improvement. While our results currently do not outperform the PS1 × WISE catalog from B22 in terms of bias or scatter, our model introduces a number of innovative features that enhance its value as a tool for specific scientific applications. Importantly, our model outputs non-parametric conditional density functions (CDFs) for photometric redshifts that allow for the full complexity of multi-modal distributions. Moreover, our model demonstrates superior probabilistic calibration compared to B22, ensuring that its predictions align more closely with the true distribution of redshifts across the test dataset. This improvement in calibration enhances the reliability of our results for follow-up analyses; for example, we believe our model could be particularly useful in providing a a prior for redshift in stellar population synthesis codes.

Another key advantage is the flexibility of our model, which allows for forced photometric redshifts on arbitrary sky coordinates without requiring a separate processing pipeline. This feature empowers users to obtain redshifts for sources outside predefined catalogs. While this shifts some responsibility to the user to consider the context of the cutouts, it also opens new possibilities for researchers requiring a more versatile tool.

Looking further to the future, the runtime of our model, as experienced through our web-app, is primarily constrained by the creation and transfer of cutouts from external services to our processing server. An obvious solution would be to collocate the cutouts services and data with our model. Ongoing efforts by STScI to host mission data on Amazon Web Services could perhaps facilitate this collocation, as we could then use AWS to host our model. We expect if such efforts materialize then our approach has the potential to become more competitive with the scaling of tabular based models today. \added{We note that both \emph{Euclid}~\citep{eulid} and LSST are adopting strategies to enable user analysis at the location of data, (see the Rubin Science Platform \citep{RubinSciencePlatform}). In these systems, we can envision deploying user-installable software packages that could be spun-up at need to perform inference. However, we expect that for our niche these platforms will of course be limited by the availability of GPUs and siloing of data from different surveys.}  

Despite current limitations, \texttt{Mantis Shrimp} adds value for researchers working with PanSTARRS photometry or requiring photometric redshifts across its footprint. Its unique capabilities position it as a useful complement to existing methods and a promising foundation for future improvements. We address specific strategies for enhancing our model to approach LS-level performance in the next subsection.

\subsection{Comparison to DESI Legacy Survey Photo-z}
\begin{figure*}[!ht]
    \centering
    \includegraphics[scale=0.55]{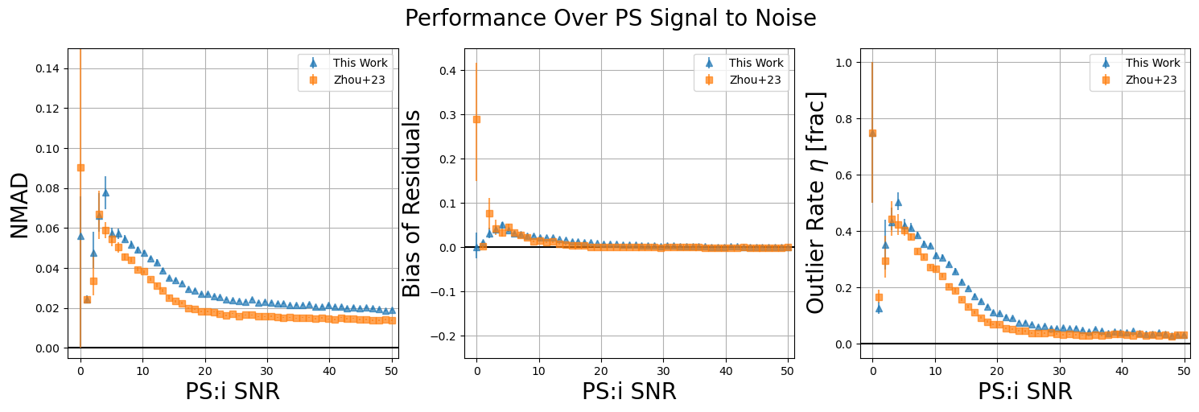} 
    \caption{\small\textbf{Explaining the Performance of Zhou+23 in the context of the limitations of the PanSTARRS survey.} In each panel we plot our performance metrics in bins of PanSTARRS SNR measured in the i-band using a Kron aperture. Error bars are determined from bootstrapping the sample within each band. This demonstrates that the model's performance is sensitive to the SNR of the underlying target galaxy, as expected. Furthermore, because the tabular Zhou+23 model outperforms this work along the breadth of SNR combined with the knowledge that the DESI Legacy Survey's photometry is overall deeper than PanSTARRS, we conclude that the quality of the underlying photometry explains why we do not match Zhou+23 performance on this sample.} 
    \label{fig:SNRanalysis}
\end{figure*}

Our comparison to \citet{Zhou_Photoz_DECaLS_DR9} shows that our computer vision models do not improve on global performance compared to tree-based tabular methods using only optical and IR bands on a very similar spectroscopic population. This is perhaps not very surprising given that we are using the PanSTARRS optical stacks while \citet{Zhou_Photoz_DECaLS_DR9} use data derived from the LS survey, which itself is about a magnitude deeper in each band and was the photometric data used for the targeting criteria of DESI. About 23\% of our dataset is not observed in PanSTARRS at the 5$\sigma$ detection level, emphasizing the inherent challenges of using PanSTARRS photometry mixed with DESI spectroscopic targeting. \added{We considered whether the underlying problem was primarily due to the low PanSTARRS SNR galaxies (which are likely still identified as objects in the LS pipeline) or whether the overall increased depth of the LS actually give a better typical SNR for each object, leading to an increase in performance of \citet{Zhou_Photoz_DECaLS_DR9} across all PanSTARRS SNR levels. This analysis is presented in Figure~\ref{fig:SNRanalysis}. There we show that in fact across a wide region of PanSTARRS SNR \citet{Zhou_Photoz_DECaLS_DR9} edges out our model, which we interpret as evidence of the importance of low noise background in utilizing both tabular and computer vision approaches to photometric redshift estimation. Future work could support this hypothesis by training a model on DESI Legacy Survey, or soon, LSST data products.}

\subsection{Interpretability}
We endeavored to explain the behavior of our neural network models with standard computer vision interpretability techniques. While our Shapley value analysis is only qualitative, we argue that certain features match onto our domain knowledge. Both early and late fusion models show a g-band MM-SHAP curve reaches an inflection point at around redshifts of $z=0.4$ corresponding to when the 4000\AA~break leaves the g-band. Shortly thereafter we observe a corresponding peak of the r-band MM-SHAP value, marking the full transition of the 4000\AA~peak into the r-band. This is a specific feature that was predicted by the physics we had introduced in section~\ref{sec:physics}. A second important observation is that both models have at least one WISE band with a positive Shapley value, which is expected given the known powerful correlation between $r - W1$ color used in the DESI selection criteria. Because the shapes of the Shapley curves are overall similar taken in tandem with the performance differences also being similar, we conclude that the choice of early vs late fusion architecture should be left to the ease-of-use considerations we have described above.

This analysis represents an exploratory foray into interpretation methods. We began this work under the premise that interpretability methods should be developed such that domain scientists can evaluate how well the model's behavior match's onto our expectations; however, in this work we have not attempted to properly quantify our expectations. Instead we have relied upon a qualitative discussion of the relevant physics to articulate how the bands importance should change as a function of redshift. More work is needed in this direction to develop rigorous frameworks of evaluating AI models for science.

The framework of targeting domain-informed features to probe the internal reasoning of computer vision models with Shapley values can be extended. For example, if we remove the flux from nearby sources but leave the targeted galaxy intact we could probe how sensitive our model is to the local environment. We could also ablate certain features within galaxies (nuclear cores, bars, spiral arms) to understand how the presence of these features affect our model. 

\added{We include in appendix~\ref{appendix:ShapleyBlends} an analysis of how blends affect our Shapley values, finding that even when masking out all objects that reasonably have a blend in any band our Shapley values do not meaningfully change. Nonetheless the rate of blends will increase with deeper surveys, so future work should consider modeling sources in images individually and ablating the flux from the target galaxy only. However, it is also worth considering the ''AI optimist'' position that neural network models can and will learn to ignore secondary objects in blends to perform well on this task. If that is true, we expect that ablating flux from the secondary object in the blend would have no meaningful impact on redshift output of the network.}

\section{Limitations and Future Work}

\subsection{Population of Spectroscopically Identified Galaxies:}
Due to the different scientific goals of individual spectroscopic surveys the population of galaxies with available spectroscopy that forms our training data is not reflective of a flux-limited sample of observable galaxies \citep{TeddyBeck2017}. As AI models often do not extrapolate well away from the training data, caution is warranted when applying our model to data outside the targeting criteria of the composite surveys used to make up our dataset. \added{We have included a first pass at characterizing the sample of confirmed spectroscopic galaxies in Figure~\ref{fig:ExplorationOptical}, but in this work have limited ourselves to colors and magnitudes only. While useful to readers who want immediate context on our sample's completeness, it is somewhat dissonant with our work's focus on utilizing neural networks for feature extraction. Future work our group has already begun will investigate characterizing the completeness and sample-density of spectroscopic galaxies utilizing features extracted from original images and spectra (Widger et. al. 2025, in prep). We note that this is similar to redshift characterization previously completed using self-organized maps \citep{Masters_2015}, but would be extended beyond color and magnitude relationships. Understanding this space is critical to understanding our model's limitations}

\subsection{Limitations of Analysis of Early vs Late Time Fusion:}
It has been pointed out in previous work that there is a significant variance on training outcomes due to the inherent randomness in PyTorch, model initialization, and random data order during training \citep{ModelVariabilityOverTraining}. When evaluating performance between two architectures, the variance inherent in training should be taken into account before making a determination of which architecture/hyperparameters are a better choice. The cost of training AI models often make this kind of analysis prohibitive. While our experience working with both early and late fusion models allows us to comment on our perceived benefits to each architecture, our experiment design prevents us from definitively saying either architecture is more performant in this task.

\subsection{Future of Surveys:}
Finally, we look forward to future public releases of both photometric and spectroscopic surveys to improve upon the \texttt{Mantis Shrimp} dataset. At time of writing, DESI is an ongoing spectroscopic mission that will add some 60 Million galaxies to our training sample, which would be a nearly 15-fold increase in targets. \added{In addition to DESI, major spectroscopic surveys have recently started or are planned which will contribute to the sample of spectroscopic galaxies. The Subaru Prime Focus Spectrograph (PFS) \citep{PFS_instrument}, started main science observations in early 2025. The PFS galaxy evolution surveys will contribute over 300k spectra, targeting an overall deeper population than DESI \citep{PFS_GE}. The Multi Object Optical and Near-infrared Spectrograph for VLT (MOONS) instrument \citep{MOONS_instrument} and its initial main extragalactic survey MOONRISE \citep{MOONRISE} will contribute about 500k galaxies focused at cosmic noon (1.0<Z<2.5) and is expected to start survey operations in October 2025. Follow-on surveys using the MOONS instrument should contribute millions more galactic spectra over the first ten years of operations \citep{MOONRISE}. Finally, the 4-metre Multi-Object Spectroscopic Telescope (4MOST) \citep{4most_instrument} will begin science operations in early 2026. The 4MOST main extragalactic surveys \citep{4MOST_cosmosurvey,4MOST_wavesurvey} will include approximately 9 million additional spectroscopically confirmed galaxies. Each of these upcoming surveys and the many auxiliary surveys associated with each instrument, have different targeting criteria, which can make understanding the exact contours of what the population of spectroscopically confirmed galaxies difficult to track.}

\added{In addition to these spectroscopic surveys, deeper photometric surveys or those with improved pixel resolution are on the horizon. The LSST \citep{LSST_main} will provide deeper optical images of the southern sky compared to the PanSTARRS survey, so would enable a southern sky variant of this work with additional u-band photometry. This improved depth would help resolve low surface brightness features of galaxies, potentially opening increasing the power of morphological features to the redshift model. Ground based IR telescopes including VISTA \citep{VISTA_main} and UKIDSS \citep{UKIDSS} could be added to our pipeline, especially if we changed our architecture to more elegantly handle masking bands of data which may be unavailable at any given pointing (e.g., borrowing ideas from the hyperspectral imaging literature \citep{HyperSpectralReview}.) Finally, upcoming major releases for space based telescopes \emph{Euclid} \citep{eulid}, \emph{SphereX} \citep{SphereX}, and the \emph{Nancy Grace Roman Space Telescope} \citep{RomanDoc} all will observe in IR photometric bands at a depth we believe would add value to our photometric redshift algorithm. \emph{Euclid} and \emph{The Roman Space Telescope} will have about twice the pixel resolution than LSST and diffraction-limited seeing, which may aid in our model's ability to utilize morphological features.}

\section{Conclusion}
We have described in detail the data preparation, pipeline, training, and analysis of the first computer vision photometric redshift estimation model for the entire northern sky. We have shown that both early and late fusion paradigms can be used to train comparable models in terms of point performance, probabilistic calibration and behavior. We show that the choice of early vs late fusion architecture are both effective in this task, so the choice should be left to the user to choose the paradigm that best fit their needs. Our computer vision pipeline is unique in that it allows users to perform a "forced photometric redshift" compared to the catalogs available from B22. While querying our web-app is as simple as supplying a RA and DEC coordinate pair, the trade-off is that we shift the responsibility to the user to evaluate the context of their own galaxy.

We make our unique dataset available to the community through PNNL's DataHub, enabling other researchers to build upon our work studying multiple instrument fusion. We make our model weights available for download, as well as our source code accompanied by Notebook style tutorials demonstrating data loading and model training. Our hope is that the availability of data and tutorial will allow our work to be a starting point for astronomers interested in deep learning.

Finally, we showed that Shapley values could be used to extract plausible explanations of the behavior of the model on the ensemble of galaxy samples. We designed a domain informed definition of null contribution of features from each channel using source extractor. This highlights the role we believe scientists play in tailoring AI methods specific to their domain. An open problem remains for how to robustly determine the physical knowledge that AI learners have extracted from datasets and phrase that knowledge in terms that humans can understand. The age of useful AI assistants for researchers is approaching, but making full use of these tools requires bridging the human-AI communication gap. 
\begin{acknowledgments}
This work would not have been possible without correspondence from Dustin Lang (Perimeter Institute for Theoretical Physics) and Aaron Meisner (NSF NOIRLab) on help with querying for UnWISE and GALEX images from the DESI Legacy Surveys. Additionally, thanks to Rick White and Travis Berger, (Space Telescope Science Institute) for help querying PanSTARRS image stacks. We thank Mike Walamsley, George Stein, Aleksandra \'Ciprijanovi\'c,
4, and A-Li Luo for correspondence regarding self-supervised and contrastive learning models. We especially thank Tim Marrinan (PNNL), Tony Chiang (ARPA-H), Annika Peter (CCAPP and OSU), Yuan-Sen Ting (CCAPP and OSU) Peter Taylor (CCAPP and OSU), and Emily Saldanha (PNNL) for providing comments and help preparing the manuscript. 

A. Engel, N. Byler, A. Tsou, E. Bonilla, and I. Smith were partially supported by an interagency agreement (IAA) between NASA and the DOE in liu of grant awarded through the NASA ROSES D.2 Astrophysics Data Analysis grant\# 80NSSC23K0474, ``Multi-Survey Photometric Redshifts with Probabilistic Output for Galaxies with 0.0 < Z < 0.6.'' PNNL is a multi-program national laboratory operated for the U.S. Department of Energy (DOE) by Battelle Memorial Institute under Contract No. DE-AC05-76RL0-1830. 
\end{acknowledgments}

%

\vspace{5mm}
\facilities{PS1, NEOWISE, GALEX, Sloan, DESI}



\label{appendix:libraries}
\software{
Astropy \citep{astropy:2013,astropy:2018,astropy:2022},
CalPIT \citep{Dey_ConditionalCalibration2022_CalPIT}, 
Dustmaps \citep{DUSTMAPs},
Einops \citep{Einops},
FFCV \citep{FFCV},
Matplotlib \citep{matplotlib},
Numpy \citep{numpy},
Pandas \citep{pandas:paper,pandas:zenodo}
PyTorch \citep{PyTorchPaszke2019},
Scipy \citep{scipy},
Source Extractor \citep{SourceExtractor}}

\appendix

\section{Facility and Institution Acknowledgment}
The Pan-STARRS1 Surveys (PS1) and the PS1 public science archive have been made possible through contributions by the Institute for Astronomy; the University of Hawaii; the Pan-STARRS Project Office; the Max-Planck Society and its participating institutes, the Max Planck Institute for Astronomy, Heidelberg and the Max Planck Institute for Extraterrestrial Physics, Garching; The Johns Hopkins University, Durham University; the University of Edinburgh; the Queen’s University Belfast; the Harvard-Smithsonian Center for Astrophysics; the Las Cumbres Observatory Global Telescope Network Incorporated; the National Central University of Taiwan; the Space Telescope Science Institute; the National Aeronautics and Space Administration under grant no. NNX08AR22G issued through the Planetary Science Division of the NASA Science Mission Directorate; the National Science Foundation grant no. AST-1238877; the University of Maryland; Eotvos Lorand University (ELTE); the Los Alamos National Laboratory; and the Gordon and Betty Moore Foundation.

The Legacy Surveys consist of three individual and complementary projects: the Dark Energy Camera Legacy Survey (DECaLS; Proposal ID \#2014B-0404; PIs: David Schlegel and Arjun Dey), the Beijing-Arizona Sky Survey (BASS; NOAO Prop. ID \#2015A-0801; PIs: Zhou Xu and Xiaohui Fan), and the Mayall z-band Legacy Survey (MzLS; Prop. ID \#2016A-0453; PI: Arjun Dey). DECaLS, BASS and MzLS together include data obtained, respectively, at the Blanco telescope, Cerro Tololo Inter-American Observatory, NSF’s NOIRLab; the Bok telescope, Steward Observatory, University of Arizona; and the Mayall telescope, Kitt Peak National Observatory, NOIRLab. Pipeline processing and analyses of the data were supported by NOIRLab and the Lawrence Berkeley National Laboratory (LBNL). The Legacy Surveys project is honored to be permitted to conduct astronomical research on Iolkam Du’ag (Kitt Peak), a mountain with particular significance to the Tohono O’odham Nation.

NOIRLab is operated by the Association of Universities for Research in Astronomy (AURA) under a cooperative agreement with the National Science Foundation. LBNL is managed by the Regents of the University of California under contract to the U.S. Department of Energy.

The Legacy Survey team makes use of data products from the Near-Earth Object Wide-field Infrared Survey Explorer (NEOWISE), which is a project of the Jet Propulsion Laboratory/California Institute of Technology. NEOWISE is funded by the National Aeronautics and Space Administration.

We gratefully acknowledge NASA’s support for construction, operation, and science analysis for the GALEX mission, developed in cooperation with the Centre National d’Etudes Spatiales of France and the Korean Ministry of Science and Technology. The grating, window, and aspheric corrector were supplied by France. We acknowledge the dedicated team of engineers, technicians, and administrative staff from JPL/Caltech, Orbital, University of California, Berkeley, Laboratoire d’Astrophysique de Marseille, and the other institutions who made this mission possible.

Funding for the DEEP2 Galaxy Redshift Survey has been provided by NSF grants AST-95-09298, AST-0071048, AST-0507428, and AST-0507483, as well as NASA LTSA grant NNG04GC89G. This research uses data from the VIMOS VLT Deep Survey, obtained from the VVDS database operated by Cesam, Laboratoire d’Astrophysique de Marseille, France. This paper uses data from the VIMOS Public Extragalactic Redshift Survey (VIPERS). VIPERS has been performed using the ESO Very Large Telescope, under the ‘Large Programme’ 182.A-0886. The participating institutions and funding agencies are listed at http://vipers.inaf.it. The WiggleZ survey acknowledges financial support from The Australian Research Council (grants DP0772084, LX0881951, and DP1093738 directly for the WiggleZ project, and grant LE0668442 for programming support), Swinburne University of Technology, The University of Queensland, the Anglo-Australian Observatory, and The Gregg Thompson Dark Energy Travel Fund at UQ. GAMA is a joint European-Australasian project based around a spectroscopic campaign using the Anglo-Australian Telescope. The GAMA input catalog is based on data taken from the Sloan Digital Sky Survey and the UKIRT Infrared Deep Sky Survey. Complementary imaging of the GAMA regions is being obtained by a number of independent survey programmes including GALEX MIS, VST KiDS, VISTA VIKING, WISE, Herschel-ATLAS, GMRT and ASKAP providing UV to radio coverage. GAMA is funded by the STFC (UK), the ARC (Australia), the AAO, and the participating institutions. The GAMA website is http://www.gama-survey.org/. This paper made use data from the Final Release of 6dFGS. the 6dFGS website is http://www-wfau.roe.ac.uk/6dFGS/.

Funding for the Sloan Digital Sky Survey IV has been provided by the Alfred P. Sloan Foundation, the U.S. Department of Energy Office of Science, and the Participating Institutions. SDSS-IV acknowledges support and resources from the Center for High-Performance Computing at the University of Utah. The SDSS website is www.sdss.org.

SDSS-IV is managed by the Astrophysical Research Consortium for the Participating Institutions of the SDSS Collaboration including the Brazilian Participation Group, the Carnegie Institution for Science, Carnegie Mellon University, the Chilean Participation Group, the French Participation Group, Harvard-Smithsonian Center for Astrophysics, Instituto de Astrofísica de Canarias, The Johns Hopkins University, Kavli Institute for the Physics and Mathematics of the Universe (IPMU) / University of Tokyo, the Korean Participation Group, Lawrence Berkeley National Laboratory, Leibniz Institut für Astrophysik Potsdam (AIP), Max-Planck-Institut für Astronomie (MPIA Heidelberg), Max-Planck-Institut für Astrophysik (MPA Garching), Max-Planck-Institut für Extraterrestrische Physik (MPE), National Astronomical Observatories of China, New Mexico State University, New York University, University of Notre Dame, Observatário Nacional / MCTI, The Ohio State University, Pennsylvania State University, Shanghai Astronomical Observatory, United Kingdom Participation Group, Universidad Nacional Autónoma de México, University of Arizona, University of Colorado Boulder, University of Oxford, University of Portsmouth, University of Utah, University of Virginia, University of Washington, University of Wisconsin, Vanderbilt University, and Yale University.

This research used data obtained with the Dark Energy Spectroscopic Instrument (DESI). DESI construction and operations is managed by the Lawrence Berkeley National Laboratory. This material is based upon work supported by the U.S. Department of Energy, Office of Science, Office of High-Energy Physics, under Contract No. DE–AC02–05CH11231, and by the National Energy Research Scientific Computing Center, a DOE Office of Science User Facility under the same contract. Additional support for DESI was provided by the U.S. National Science Foundation (NSF), Division of Astronomical Sciences under Contract No. AST-0950945 to the NSF’s National Optical-Infrared Astronomy Research Laboratory; the Science and Technology Facilities Council of the United Kingdom; the Gordon and Betty Moore Foundation; the Heising-Simons Foundation; the French Alternative Energies and Atomic Energy Commission (CEA); the National Council of Science and Technology of Mexico (CONACYT); the Ministry of Science and Innovation of Spain (MICINN), and by the DESI Member Institutions: www.desi.lbl.gov/collaborating-institutions. The DESI collaboration is honored to be permitted to conduct scientific research on Iolkam Du’ag (Kitt Peak), a mountain with particular significance to the Tohono O’odham Nation. Any opinions, findings, and conclusions or recommendations expressed in this material are those of the author(s) and do not necessarily reflect the views of the U.S. National Science Foundation, the U.S. Department of Energy, or any of the listed funding agencies.
 
The Planck dust map is based on observations obtained with Planck (http://www.esa.int/Planck), an ESA science mission with instruments and contributions directly funded by ESA Member States, NASA, and Canada.

\section{Recalibration of conditional density estimates using CalPIT Library}
\label{appendix:CalPIT}

In this appendix, we discuss our effort to re-calibrate the \texttt{Mantis Shrimp} model using the \texttt{CalPIT}
CalPIT trains an auxillary neural network to learn a local calibration to our model's conditional density estimate. Such local calibration are an important recent improvement over global calibration (e.g., see \citep{GlobalCalibrationUsingPIT_Bordoloi2010}), since it had been shown that different sub-populations can adversarially combine together to offset regions of the global PIT visualization \citep{DESI_redshifts_Zhou21}. 

CalPIT takes as input a feature vector, which in our case must be extracted from the images. We use the \texttt{sep} library to measure Kron magnitudes in each band. Just as in our Shapley Value baseline image pipeline, if no 2$\sigma$ sources are detected within 5 pixels of the center of the image, we instead measure forced photometry using a static circular aperture of radius 4 pixels centered in the cutout. From these Kron magnitudes, we calculate a feature vector that includes all pairwise colors from each of the nine bands, the magnitudes themselves, and the extinction along line of sight, for a total of 84 unique features.

We follow many of the suggestions of the CalPIT authors: we choose as our architecture a 3-layer monotonically increasing neural network \citep{Monotonic_NNs} with hidden width=1024, learning rate=1e-3, and weight decay=1e-5. We diverge from the original authors by the \texttt{Mantis Shrimp} validation and computed CDEs as the training data for CalPIT. We do this out of an abundance of caution, reasoning that it is possible for the \texttt{Mantis Shrimp} model to have overfit the CDEs onto training data in a way that would make the model acting on the sub-population of our exact training data different than when acting on unseen validation data. (We know that the \texttt{Mantis Shrimp} model does behave differently on training data given the training-data vs validation-data loss curves). We train CalPIT for 100 epochs and evaluate on the \texttt{Mantis Shrimp} test dataset. This choice allows us to compare CDE losses directly before and after re-calibration.

\begin{figure}[!ht]
    \centering
    \includegraphics[scale=0.6]{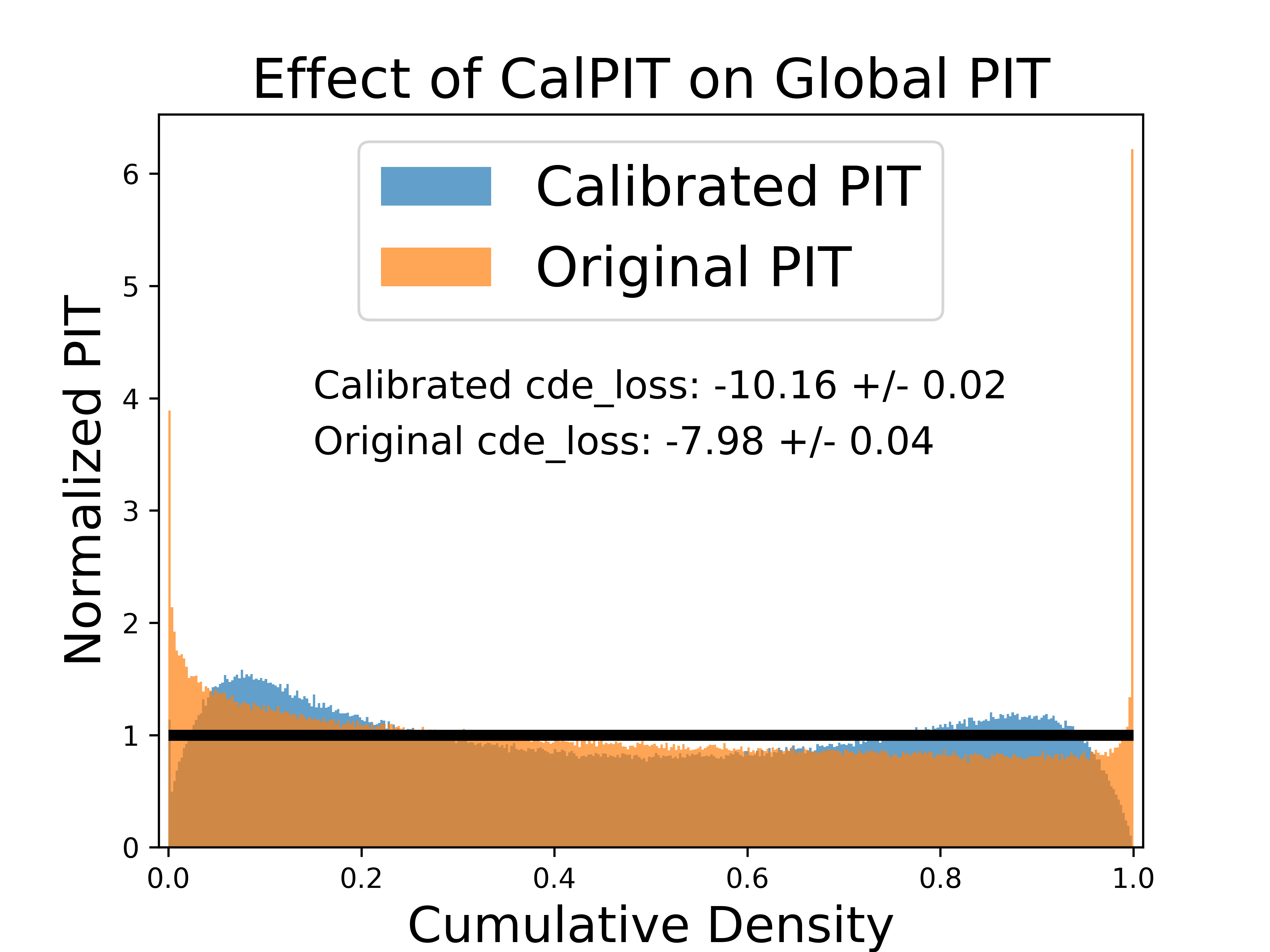} 
    \caption{\small \textbf{Effect of \texttt{CalPIT} on Global PIT} We see how CalPIT recalibrates the model to reduce the amount of catastrophic outliers, but in doing so reduces their rate below what we should expect. We nonetheless believe this calibration preferable given that we see a major use case of our model as providing a prior over redshift for downstream analysis; in which case a less informed prior would be preferred over assigning zero probability to the true value.}
    \label{fig:CalPIT_PIT}
\end{figure}

We show the global calibration effect of CalPIT in Figure~\ref{fig:CalPIT_PIT}, computing that the CDE loss improves from the original value -7.98 to a calibrated value -10.16, which demonstrates that overall CalPIT calibration achieves a better overall probabilistic calibration. The PIT visual metric confirms this, showing that many probabilistic catastrophic errors are removed (i.e, the bins on the far left and right side of the PIT), moreover, that the re-calibration makes the rate of probabilistic outliers go to nearly 0. While this is certainly not the ideal behavior, we do consider that having a generally high chance the CDEs we generate will have coverage over the true value a better outcome than too-tight CDEs which may assign a very small, or even zero probability to the true value. For example, one user-group of our model will be those who want to use our CDEs as an initial prior for more rigorous redshift modeling, e.g., MCMC methods for stellar population synthesis such as the prospector library \citep{Prospector}.

\begin{figure}[!ht]
    \centering
    \includegraphics[scale=0.8]{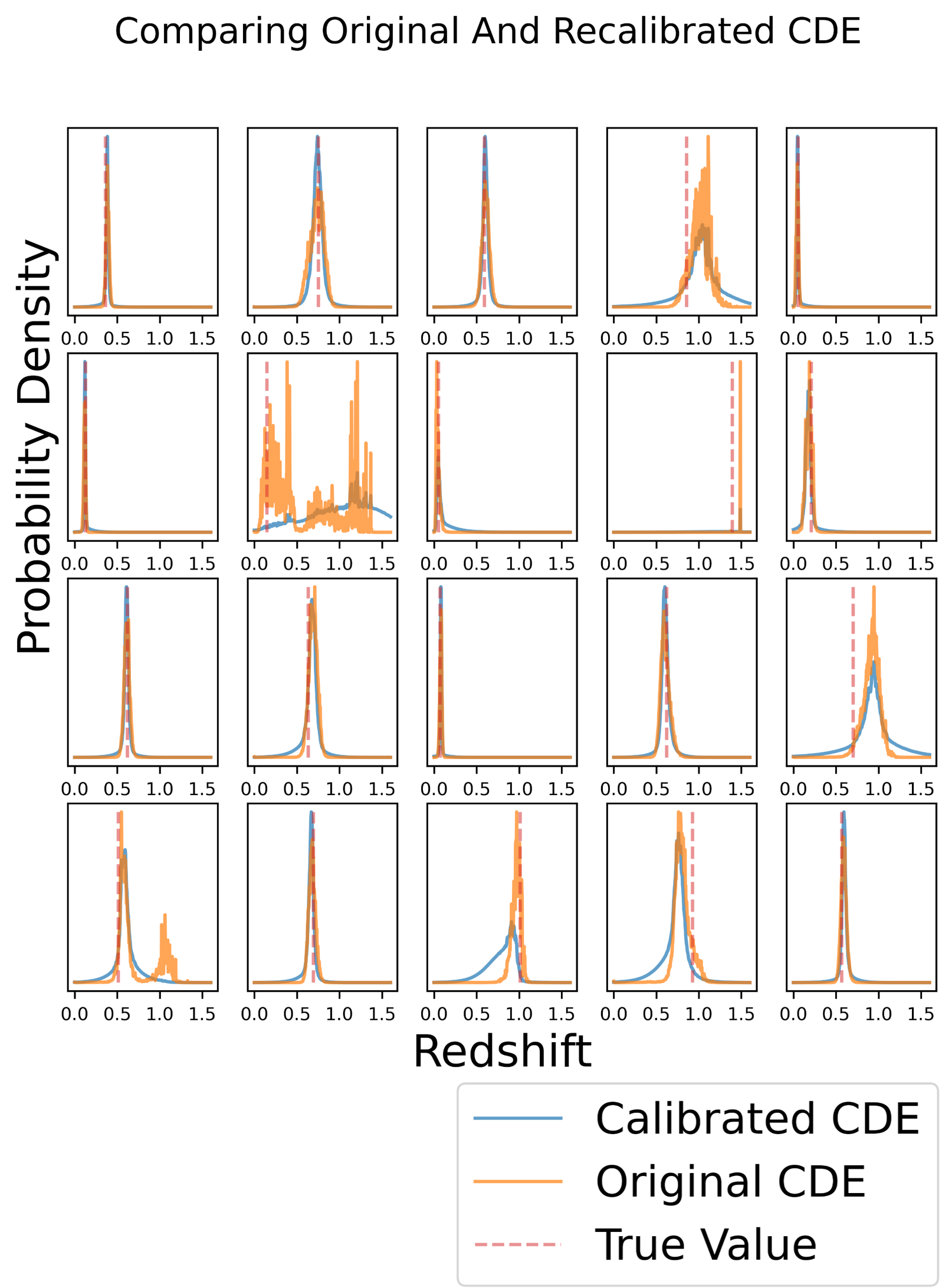} 
    \caption{\small \textbf{A set of examples of CDE estimates before and after CalPIT calibration}. While we have observed samples where CalPIT can remove bi-modality, this random sample of our original and calibrated CDEs show that CalPIT more often does not radically change the initial estimate.}
    \label{fig:CalPIT_cde_sample}
\end{figure}

Of course, reducing the rate of probabalistic catastrophic errors could be achieved by monstrously widening the CDE estimates; however, we evaluated by eye and do not observe CalPIT monstrougly widening the CDEs. We plot a sample of the calibrated and original CDE estimates in Figure~\ref{fig:CalPIT_cde_sample}, and the reader may verify that overall the CalPIT re-calibration and the original CDEs share a similar shape, as is be expected. CalPIT does not simply widen or tighten the CDE, but can actually reshape it entirely-- meaning it affects the point estimate performance of our model. We do note that at times we observe CalPIT removing bi-modality from an initial CDE, and though we did not observe it on our sample, the original authors note that CalPIT can also insert bi-modality into a CDE.

As mentioned in the previous paragraph, CalPIT can alter the point-prediction we infer from the expectation value of the calibrated CDE. We next evaluated how this point performance changed. In Table~\ref{tab:calpit_pointestimate} we report before and after calibration point estimate performance using the metrics defined in Section~\ref{sec:metrics}. We find that the calibration generally trades-off a lower variance for a lower bias, (now being competitive in terms of Bias with B22 on the PanSTARRS $\times$ WISE galaxies). Because users may desire a lower NMAD or lower Bias depending on their specific use case \citep{PZEvaluationLSSTSchmidt2020}, we leave the option of using CalPIT to the user on our web-app. 

\begin{table*}[!ht]
\caption{Effects of CalPIT calibration on point estimate performance. Overall while re-calibration does improve the probability estimate calibration it also predictably increases scatter (this is predictable given that our initial estimates were overall too confident); it unfortunately also increase Bias and outlier rate which is not necessarily expected. Future studies should investigate either alternative inputs to CalPIT, e.g., providing the entire input cutouts to CalPIT, or software alternatives. We leave using the CalPIT re-calibration as an option on the web-app.}
\centering
\label{tab:calpit_pointestimate}
\begin{tabular}{ccc}
    \toprule
    Metric & Before Calibration & After Calibration \\ 
    \hline
    NMAD ($\times 10^{-2}$) & 2.438(4) & 2.82 \\
    Bias ($\times 10^{-2}$)& 1.151(9) & 5.74 \\
    $\eta$ (\%) & 17.79(4) & 21.50 \\
    CDE loss & -7.45(7) & -10.16(2) \\
     \hline
    \bottomrule
  \end{tabular}
\end{table*}

Future work in the direction of improving our calibration specifically using CalPIT would be to find more rigorous means of evaluating the claim that CalPIT achieves a local feature calibration. We consider this evaluation beyond the scope of our work, but note that the solution would need to consider the full joint-structure of all features. Regarding our own usage of CalPIT, future work could investigate using AI models to extract highly informative feature vectors from cutout images. We refrain from using our mantis-shrimp model as a feature encoder as prior work has shown how classification models may collapse-out unnecessary feature information from embedding vectors to perform their task \citep{neuralcollapse}, i.e., we believe we would need a model trained to be a feature-encoder model rather than a highly specialized redshift model.

\section{Expanded set of metrics}
\label{appendix:eta015}
\added{In this section we report the expanded set of metrics for the point performance of our model, notably including \textbf{$\eta_{>0.15}$}. We chose to only report our \textbf{$\eta$} (\textbf{$\eta_{>0.05}$}) in the main body to simplify the plot as most authors we compare to only report \textbf{$\eta_{>0.05}$} despite \textbf{$\eta_{>0.15}$} being overall more popular. We report the results in Table~\ref{tab:etabig}.}

\begin{table*}[!ht]
\caption{\small\textbf{Model Performance Across Sub-populations:} Mostly the same as Table~\ref{tab:subpop_reults}, but including \textbf{$\eta_{>0.15}$} and focused on only our model's results. \textbf{Values reported with leading digit of uncertainty as parenthetical}}
  \centering
 \label{tab:etabig}
  \begin{tabular}{llccccccc}
    \toprule
    Subpop. & Model & NMAD ($\times 10^{-2}$) & Bias ($\times 10^{-3}$) & $\eta$ (\%) & \textbf{$\eta_{>0.15}$} (\%) \\
    \hline
    \multirow{2}{*}{SDSS MGS} & Multi-instrument CNN, early (This Work) & 0.995(4) & -0.19(4) & 0.53(2) & 0.0007(3)  \\
    & Multi-instrument CNN, late (This Work) & 1.023(4) & -0.50(4) & 0.59(2) & 0.0007(3) \\
     \hline
     \multirow{2}{*}{PS1 $\times$ WISE}  & Multi-instrument CNN, early (This Work) & 1.862(3) & 2.88(5) & 6.59(3) & 0.61(1) \\
     & Multi-instrument CNN, late (This Work) & 1.882(3) & 3.46(5) & 6.74(3) & 0.63(1)  \\
     \hline
     \multirow{2}{*}{PS1 Objects}  & Multi-instrument CNN, early (This Work) & 1.964(3) & 4.72(5) & 8.72(3) & 4.51(2) \\
     & Multi-instrument CNN, late (This Work) & 1.989(3) & 5.64(6) & 9.00(4) & 4.51(2)  \\
     \hline
     \multirow{2}{*}{DESI LS DR9} & Multi-instrument CNN, early (This Work) & 1.960(5) & 5.43(7) & 7.93(5) & 0.85(2) \\
      & Multi-instrument CNN, late (This Work) & 1.977(5) & 6.04(7) & 8.18(5) & 0.89(2)  \\
     \hline
    \bottomrule
  \end{tabular}
\end{table*}

\section{Spectroscopic Survey Descriptions}
\label{appendix:spectroscopic}

\subsection{SDSS}
The vast majority of SDSS spectroscopy comes from three surveys with different targeting guidelines and science goals. The main galactic survey's goal was to provide uniform spectroscopy and ugriz photometry of extragalactic targets in the northern galactic cap \citep{SDSS_DR8}. It targets galaxies brighter than $r=17.77$ (80\% of the survey), Luminous Red Galaxies (LRGs) \citep{SDSS_LRG_definition}, and Quasi-stellar objects (QSOs). The main sample of galaxies brighter than $r=17.77$ is a flux limited sample \citep{SDSS_MGS}, while the LRGs are a volume limited sample designed to probe a specific sub-population of galaxies \citep{SDSS_LRG_definition}, and have a slightly more complicated targeting criteria. There is a strong correlation between satisfying LRG criteria and being an ``early-type'' or Elliptical galaxy.

The BOSS \citep{BOSS_target_guidelines} survey's goals were to vastly expand the number of available galaxies with spectroscopic redshift to perform baryonic acoustic oscillation analysis \citep{BAO}. At low redshifts, BOSS primarily targets LRGs for their strong spectral features allowing precise redshift measurement without long observation time, and beyond redshifts of 0.6 targets a stellar-mass limited sample \citep{BOSS_target_guidelines}. Finally, the extended BOSS (eBOSS) survey \citep{EBOSS} targets LRGs focused at higher redshifts than achieved with BOSS. The effect of the SDSS targeting from BOSS and eBOSS are a sample bias towards the most luminous, red galaxies. While this makes up the vast majority of the sample, SDSS also makes public auxillary programs and surveys completed at a much smaller scale. Nonetheless, it is believed that the sample bias severely restricts empirical photometric redshift algorithms in feature space outside of the flux-limited MGS.

We queried SDSS CAS jobs for all objects in SDSS north of dec>-30$^\circ$ in increments of a few ten degrees.  We accept redshifts that have no bits set on \texttt{ZWARNING}, meaning nothing is ``wrong'' with the redshift measurement. We also make a cut on the respective spectroscopic class being GALAXY. We also query \texttt{z\_noqso} and \texttt{z\_person}, which are additional fields in \texttt{SpecObjALL}. If \texttt{z\_person} is set, we accept \texttt{z\_person} instead. If \texttt{z\_noqso} is set, we accept \texttt{z\_noqso} instead. If neither \texttt{z\_person} or \texttt{z\_noqso} is set (majority of the sample) we accept the standard reported redshift of the target.  

\subsection{DESI}
DESI targets four sub-populations of galaxies, with which by our cut to ``galaxies'' we are concerned with the bright galaxy sample (BGS), the LRG sample, and the emission line galaxy (ELG) sample. We document the kinds of galaxies and at what redshifts DESI targets them below, and then describe the sample included in the DESI early data release used in this article. The BGS will include a magnitude-limited sample to $r<19.5$, which will extend our representative sample of galaxies to redshifts of approximately $z=0.6$ \citep{DESI_BGS}. The LRG sample builds upon the works of the SDSS surveys, targeting LRGs in the redshift region of $0.4<z<0.8$ in a higher density than SDSS \citep{DESI_LRG}. The ELG survey, traces the highest regime of star-forming galaxies at the peak of the highest star-forming times in the Universe, and will observe targets in $0.6<z<1.6$, specifically focusing on the region $1.1<z<1.6$ \citep{DESI_ELG}. 

These surveys are on-going and at time of writing only the DESI early data release (EDR) was publicly available \citep{DESI_EDAspectroscopy}. The DESI EDR was used as a science validation sample for the targeting and quality of the DESI surveys, and includes 430,000 bright galaxies, 230,000 LRGs, and 440,000 ELGs. \added{DESI DR1 has since become publicly available \cite{DESIDR1}}

We downloaded all the spectroscopic targets from the early data release of the DESI \citep{DESI_EDAspectroscopy}. We made quality cuts on ZWARN=0, and spectroscopically confirmed as a galaxy. We also cut at the PanSTARRS field so only accept targets above -30$^\circ$. Finally we drop all second or higher references to the same photometric object (as given by photoObjID\_survey).

\subsubsection{DEEP2}
The DEEP2 survey was designed to probe galaxy evolution and the large scale structure with sufficient high quality redshift determinations specifically for fainter objects at greater redshifts \citep{DEEP2_design}. It forms a magnitude-limited sample above redshifts of 0.75.
We download the entire DEEP2 \citep{DEEP2} catalog and make cuts on declination greater than or equal to -30$\circ$, type of best-fitting template = galaxy, and redshift quality code either secure or very secure. We record zBest for the redshift of the remaining set of galaxies.

\subsection{GAMA}
The GAMA survey target galaxies to r-band magnitude  $16.6<r<19.8$ with aims to study the dark matter halo mass function, the star formation efficiency of galaxies, and to make a measurement of the recent galaxy merger rate \citep{GAMA_DR2}.
We download all the spectroscopic targets (all galaxies for GAMA) and accept objects with photometric redshift quality=3 or quality=4 (which should amount to a 95\% probability or confidence in the photometric redshift). We drop objects with repeated CATID, and filter out objects with a declination less than $30^\circ$.

\subsubsection{VVDS}
The VVDS survey's goals were to compliment the SDSS and 2dF \citep{2dF_finalrelease} survey probing galaxy evolution by surveying galaxies at higher redshifts than those works with a simple selection function that $I_{AB}$<24mag.  
We download the final data release of the VIMOS VLT Deep Survey (VVDS) \citep{VVDS_final_release} which includes the catalogs of the individual WIDE, DEEP and Ultra-DEEP VIMOS surveys. We make a cut on the zflag table to values of 3 or 4. 

\subsection{VIPERS}
The goal of the VIPERS survey \citep{VIPERS_final} was to construct a complimentary survey at greater depths than SDSS, but over a large volume of sky. The selection functions were designed to focus the survey on the redshifts greater than 0.5 and to a depth of $i_{AB} < 22.5$ mag. 
We download the entire VIPERS catalog and make cuts on declineation greater than or equal to -30$^\circ$, that the reported zflg quality column indicates either a high confidence, highly secure, or very secure (indicating at least an estimated 90\% confidence in the redshift values). We make a cut on classFlag to only accept main VIPERS galaxy targets.

\subsection{6DF}
The goal of the 6dF \citep{6dFGS_final} survey was to provide a survey of nearby galaxies over a larger volume of the sky than contemporarily available in the preceding 2dF \citep{2dF_finalrelease} or SDSS surveys. The survey centers at a median redshift of 0.05 with a designed K-band limit 12.65. 
We downloaded the 6dF Galaxy Survey Redshift catalog Data Release 3, which contains spectroscopy, redshifts, and peculiar velocities for a selected sample of near-infrared observations primarily compiled from the 2MASS catalog. The 6dF catalog includes observations from other surveys including SDSS, 2dFGRS, and ZCAT that satisfy their target constraints. We make a cut on the quality column and accept values of 4 or 3.

\subsection{WiggleZ}
The WiggleZ survey was designed to probe large scale structure and dark energy \citep{WiggleZ_final}. The sample is selected by UV brightness using the GALEX space telescope ($\text{NUV} < 22.8$ mag). Additional targeting criteria were used to specifically target ELG. We download the complete WiggleZ catalog and make cuts on the reported quality of the redshift measurements, QUALITY = 4 or QUALITY=5, 5 meaning an excellent redshift with high S/N that may be suitable as a template. 4 meaning a redshift that has multiple (obvious) emission lines all in agreement. We also make our cut on the declination of the target being greater than or equal to -30$^\circ$.

\section{Photometric Data Collection}
\label{appendix:photometry_processing}
\label{appendix:photometry_download_and_description}
Photometric Data can be accessed from server cutout API tools.
\footnote{GALEX Cutouts: \url{www.legacysurvey.org/viewer/fits-cutout?ra=RA\&dec=DEC\&size=32\&\\pixscale=1.5\&layer=galex}}
\footnote{PanSTARRS Cutouts: \url{https://ps1images.stsci.edu/cgi-bin/fitscut.cgi?size=170\&format=fits\\\&ra=RA\&dec=DEC\&red=/rings.v3.skycell/1063/090/rings.v3.skycell.1063.090.stk.g.unconv.fits}}
\footnote{UnWISE Cutouts: \url{www.legacysurvey.org/viewer/fits-cutout?ra=RA\&dec=DEC\&size=32\&\\pixscale=2.75\&layer=unwise-neo7}} 
We query public APIs of image cutout servers from each photometric survey source at the target coordinates. This process can be easily automated with standard wget tools, but data collection is throttled by server response times and one must take care not to overload servers with requests. Data collection of our entire dataset takes 35 days, for a total volume of approximately 1.5 TB for 4.2e6 samples after quality cuts and various losses. Downloaded files are in the FITS \citep{FITS} file standard. Images are downloaded so that a 30 arcsecond per-side cutout can be created from rotated images without interpolation, with a minimum of 32pix/side as some convolutional architectures would otherwise fail. We downloaded PanSTARRS at 170pix/side, GALEX at 32pix/side, and UnWISE at 32pix/side. We use UnWISE DR7 \citep{7_yr_UnWISE}.

After downloading the cutout files we inspect them for masked pixel values. The PanSTARRS data contains masked pixels with NaN values. \added{In a sample of 50k galaxies, we observe that only 0.02\% have any masked pixels intersecting the centered target galaxy in the PanSTARRS optical stacks.} We nonetheless filter out PanSTARRS images that have more than 1\% NaN pixels in any band, then fill NaN values with pixel value 0. \added{The choice to infill NaN values is valid but could be improved, for example, since codifying our pipeline advanced methods for infilling have emerged \cite{infilling}.} This pre-processing step allows us to keep images containing small numbers of NaNs inside our input volume. GALEX and UnWISE were not observed to contain any NaN pixels, though GALEX does contain 0-valued pixels wherever a patch of sky is queried where observations are not taken.

\section{Preparing Comparison Datasets}
\label{appendix:comparison_details}



In addition to reporting our results on the held-out test dataset, we compare our analysis with the photometric redshift estimates from two available catalogs, the \citep{Pasquet2019} (P19) catalog created from the SDSS spectroscopy (including BOSS and EBOSS) of galaxies with a dust corrected SDSS Petrosian magnitude $r < 17.77$ (mimicking the SDSS main galactic sample's photometric cutoff \citep{SDSS_MGS}), and the B22 catalog (otherwise known as the WISE-PS1-STRM \citep{WISE-PS1-STRMBeck22} catalog). 

The P19 catalog differs from the original in that we query all available spectroscopic targets that satisfy SDSS Petrosian Magnitude $r < 17.77$ in SDSS DR18 \citep{SDSS_DR18}, the most recent available release at time of writing. P19 was queried from SDSS DR12 \citep{SDSS_DR12}, which was at the time then the most recent available release.  We complete the query on the SDSS CAS jobs server.
We simultaneously query for the B16 \citep{Beck2016LLNSDSS} photometric redshifts calculated using a local linear regression as a second benchmark on this dataset. To form the sample of SDSS MGS we report on, we join the CAS server query results to datapoints that had been randomly sorted into our test dataset from our original train-val-test split. In our test dataset this leaves 7e3 number of samples to evaluate upon. In principle, it would be appropriate to increase this sample by performing a catalog search between all spectroscopic targets from our larger survey in the SDSS footprint that satisfy $r < 17.77$.

A second relevant dataset that dovetails closely with the main idea of this work, combining multiple surveys together for photometric redshift information, is \citet{Beck21PS1STRM} and \citet{WISE-PS1-STRMBeck22} (B22), the latter creating the WISE-PS1-STRM catalog. While a full description can be found by reading over these works, briefly: our datasets differ in a few ways. 
First, Beck's work utilizes dense networks on pre-calculated photometric features available in catalogs while our work uses the science images. This generally means that Beck's work relies on a detection in at-least one band of each PanSTARRS and WISE for any spectroscopic target. In contrast, our work is more akin to forced photometry centered on spectroscopic astrometry-- we do not require any detections in the underlying photometric survey catalogs. 
Beck also employs a cautious astrometry check between objects in catalogs to en-masse join the catalogs and spectroscopic targets together\citep{CrossMatchBudavariSzalay}. This is entirely circumvented in our methodology since we do not need to query for objects from catalogs, but instead broad patches of the sky from image cutout servers.
B22 does not include the DESI spectroscopic target contribution, as their work was completed before the DESI early release was available to the public. Given that the DESI is nearly a quarter of our sample, we gain a sizable increase in total available targets to train on.
Finally, the WISE-PS1-STRM algorithm actually is two separate AI algorithms: the algorithm performs STAR-GALAXY-QSO filtering, and then on the objects identified as galaxies, will compute a photometric redshift. This acts as another filter when querying the WISE-PS1-STRM dataset compared to our original training sample.

\section{Additional Examples of Null Feature / Baseline Images using sep}
\label{appendix:sep_examples}
\begin{figure}[!ht]
    \centering
    \includegraphics[scale=0.45]{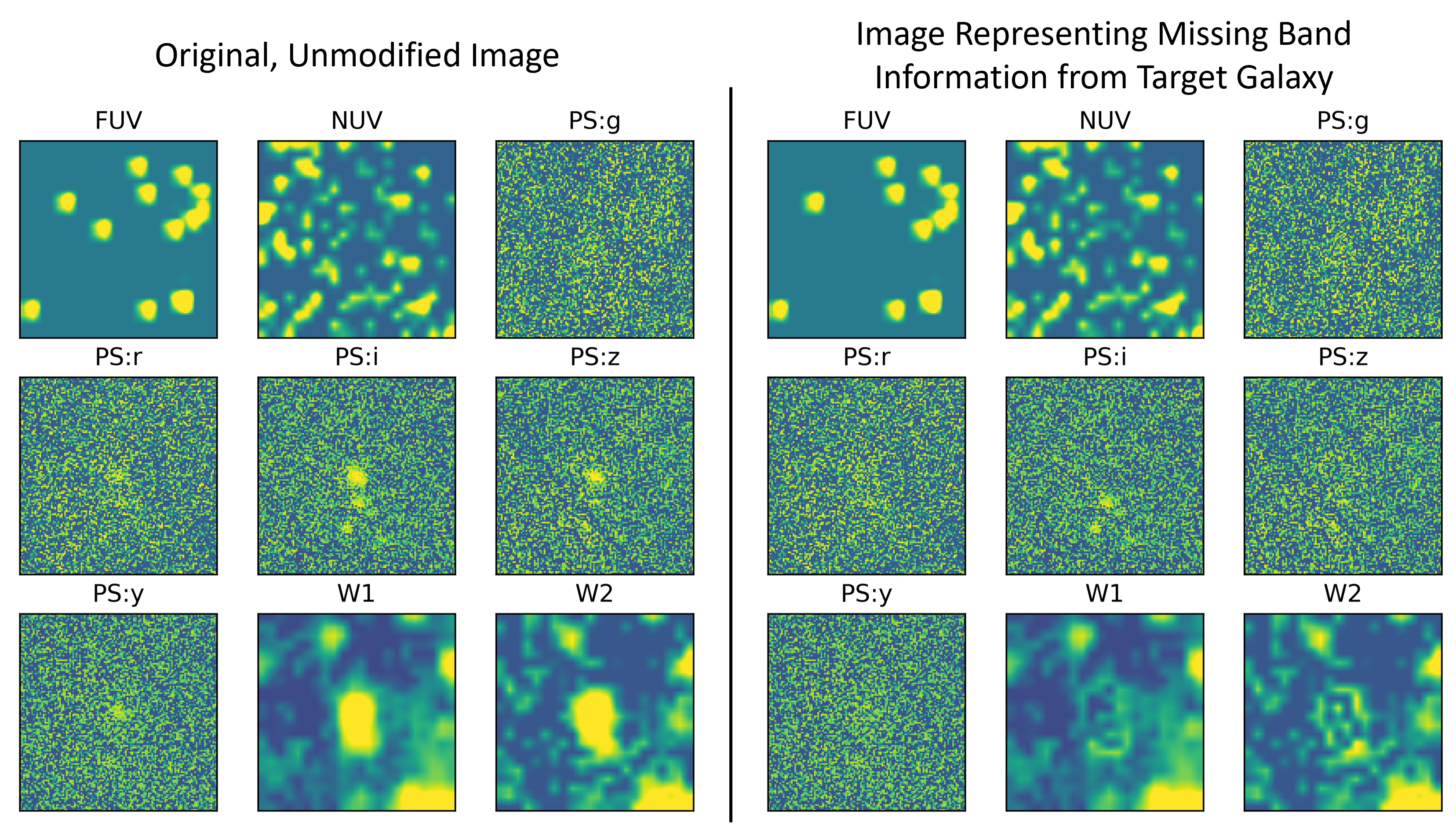}
    \caption{Example of Null Feature Contribution}
    \label{fig:null_feature_1}
\end{figure}

\begin{figure}[!ht]
    \centering
    \includegraphics[scale=0.45]{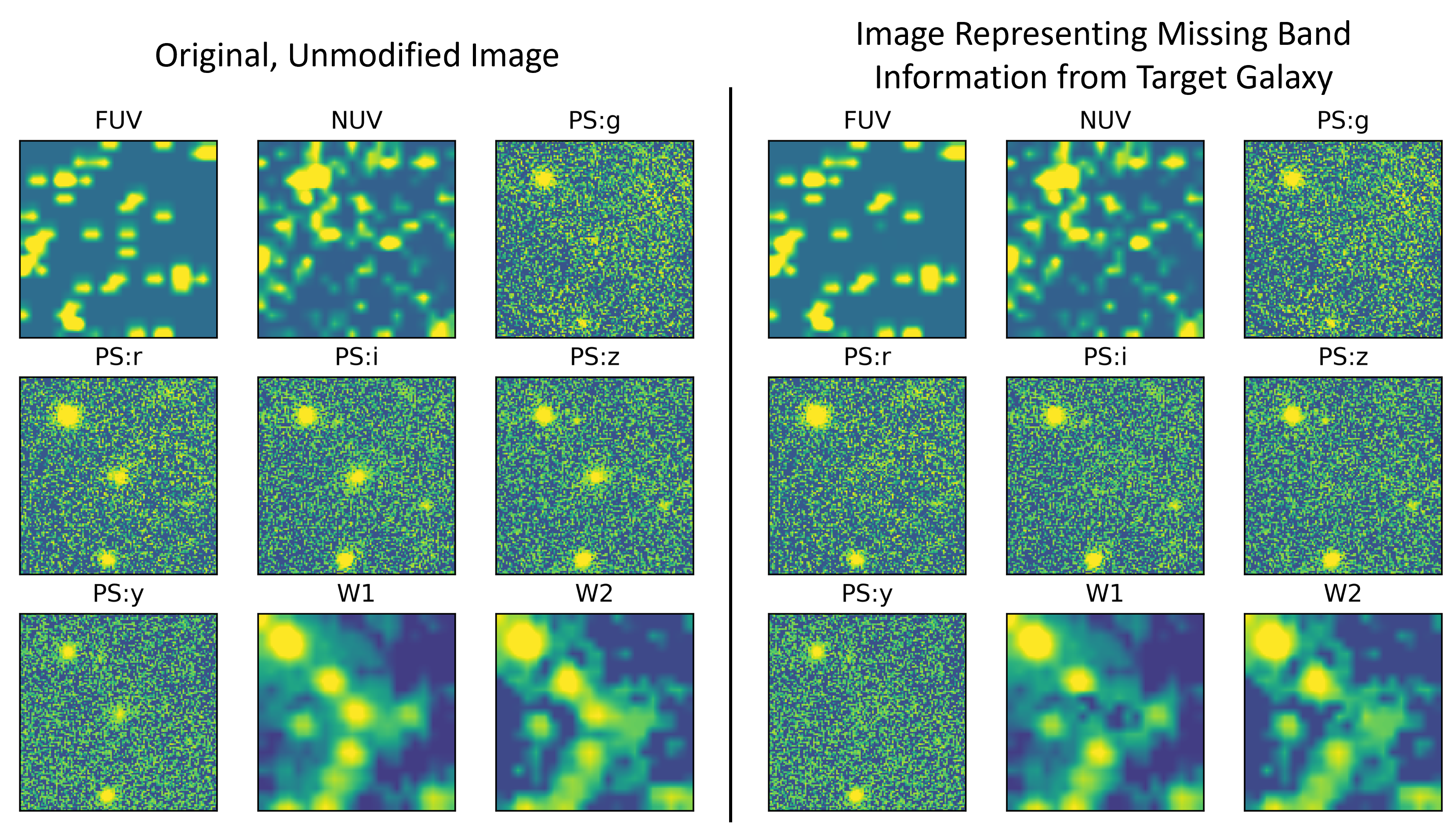}
    \caption{Example of Null Feature Contribution}
    \label{fig:null_feature_2}
\end{figure}

\begin{figure}[!ht]
    \centering
    \includegraphics[scale=0.45]{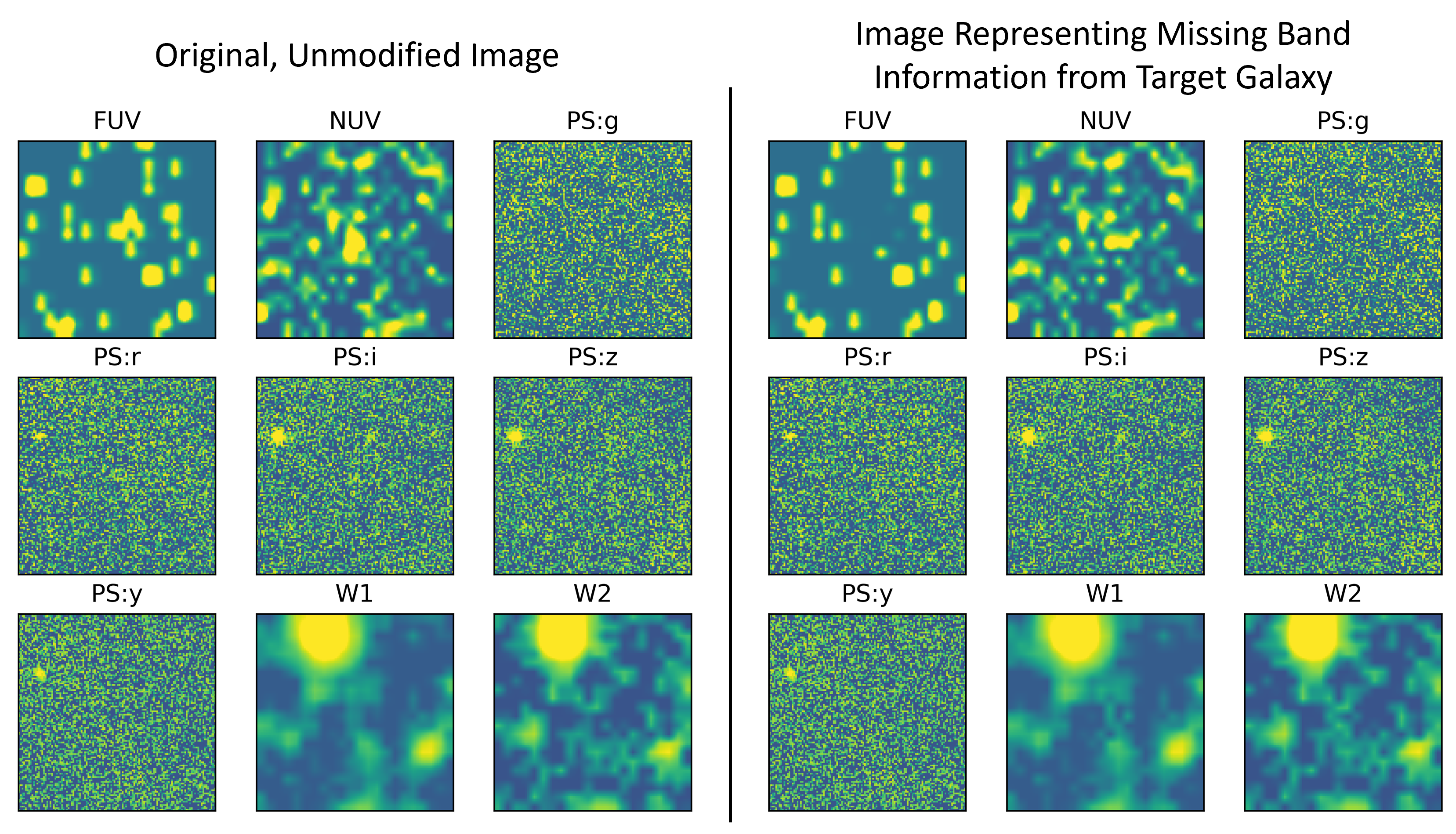}
    \caption{Example of Null Feature Contribution}
    \label{fig:null_feature_3}
\end{figure}

In this section we provide additional examples of our null feature contribution definition used in our Shapley value computation (Figure~\ref{fig:null_feature_1}, Figure~\ref{fig:null_feature_2}, and Figure~\ref{fig:null_feature_3}). In each graph, the left side shows the original images from each band, and the right shows the image representing our feature ablation. As explained in the main text, the intuition for our method is that only the difference in the NN output between the original and the feature-ablated images contributes to the Shapley value \eqref{eq:shapleyvalue}, so to target explainability to the flux of the target galaxy we ablate flux from that central source. The examples provided below are chosen at random from our dataset to provide the reader with some additional context of the current state of our implementation. We note that currently we are imperfect about removing all structure from the center of the images (which could conceivably be improved with more precise photometric modeling) and did not account for blending effects. 

\section{Evaluating the Effect of Blends on Shapley Values}
\label{appendix:ShapleyBlends}

\begin{figure}[!ht]
    \centering
    \includegraphics[scale=0.45]{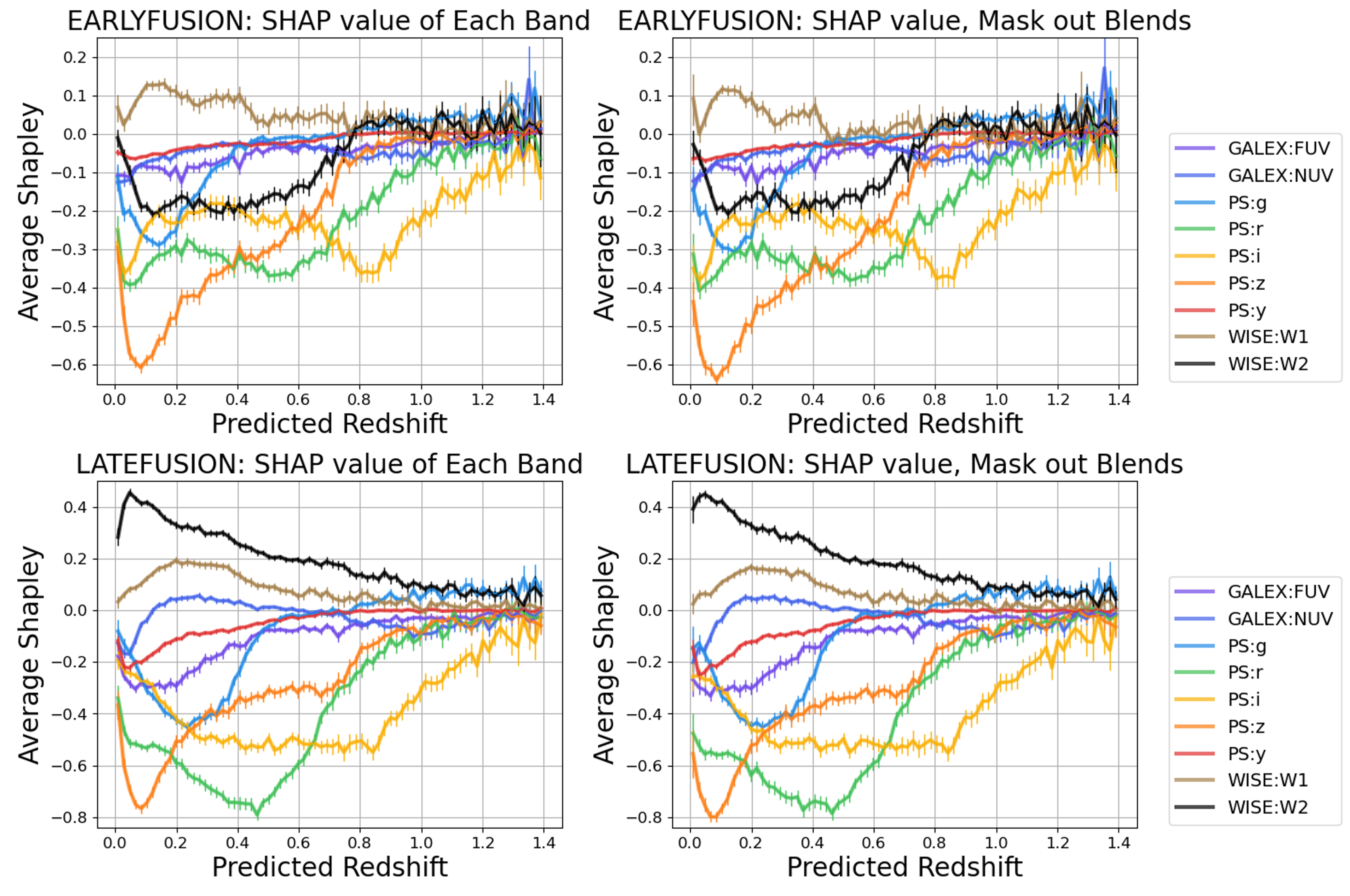}
    \caption{for both early (top row) and late fusion (bottom row) we recompute the Shapley values while excluding from the computation all images containing a blend in any band (right side). On the left side of the plot, we show again the Shapley values from the main body. There is no substantial change in the patterns of our plot, thus, we declare that our Shapley values are robust against blended objects. Recall that MM-SHAP values are just a normalization of the Shapley values, so if there is no substantial change in the Shapley values, there will be no change in the relative importance of each band as well.}
    \label{fig:BlendsAppendix}
\end{figure}

\added{In this section we detail our effort to ascertain the effect that blends have on our Shapley values, concluding that for our specific models blends do not meaningfully sway the Shapley Values away from those presented in the main body of this work. The main result is depicted in Figure~\ref{fig:BlendsAppendix}, where we mask out all objects from our Shapley computation that contain a blend in any band (26\% of objects). In the next paragraph we describe how we identified whether a band contained a blend of the target galaxy.}

\added{To determine whether the target galaxy in an image is blended with another object, we first run our object detection pipeline with source extractor and create a segmentation mask for each object in the image based on the elliptical profile of the object. Recall that in each of our cutouts we expect the target galaxy to be the central source in the image, since we queried the cutout services at the spectroscopic survey's reported astrometry. Using the same procedures as described in the main text, we accept the most central source within 5-pixels distance as the target object. Using it's segmentation mask and the masks of all other objects, we compute the percentage of the target source's pixels that overlap with another object. We define that any central source with more than 5\% of its pixels also being shared with another to be a blend. We apply this criteria to each blend in the sample used for our Shapley calculation and find the percentage of images without blends in each band to be: GALEX: FUV 0.983, GALEX: NUV 0.950, PS:g 0.946, PS:r 0.907, PS:i 0.869, PS:z 0.911, PS:y 0.951, WISE:w1 0.936, and WISE:w2 0.968. To create the comparison shown in Figure \ref{fig:BlendsAppendix}, we mask out any image in which any band contains a blend.}

\section{\emph{Mantis Shrimp} Name Inspiration}
\label{appendix:name_inspiration}
\added{In presenting this work the first question we often receive is what was the inspiration for naming our project after the Mantis Shrimp? We would like to take the opportunity to answer fully. The mantis shrimp (the common name given to many stomatopod species) is a remarkable creature in many respects, but important for this context are the unique eyes it possesses. Some species of mantis shrimp can have between 16 to 21 different photoreceptors \citep{MantisShrimpOpticalReview}, the most of any animal known. The \emph{haptosquilla trispinosa} species specifically was evaluated to understand its color vision capability in \citet{MantisShrimpSystem}, directly observing 11 different effective band-pass regions spanning the UV and Optical spectrum (with evidence to support the 12$^{\text{th}}$ expected band). The mantis shrimp optical system is therefore in great analogue to the focus of our work, which is to combine nine band-pass filters spanning from the UV to the IR together to essentially allow our network to break down color degeneracies otherwise present in the optical bands only.}

\pagebreak

\bibliography{bibliography}{}
\bibliographystyle{aasjournalv7}

\end{document}